\documentclass[12pt]{article}
\pdfoutput=1
\usepackage{latexsym}\usepackage{epsfig,amssymb,euscript}
\usepackage{amsmath} \usepackage{bbm,slashed}
\usepackage[usenames,dvipsnames]{xcolor}
\usepackage{ulem}
\usepackage{cite}
\usepackage{graphicx}
\usepackage{hyperref}

\newcommand{\bea}{\begin{eqnarray}}
\newcommand{\eea}{\end{eqnarray}}

\newcommand{\nn} {\nonumber}

\newcommand{\pa}{\partial}

\newcommand{\T}{\text{T}^{1,1}}

\newcommand{\tx}{\text}

\newcommand{\del}{\partial}

\newcommand{\bi}{\begin{itemize}}
\newcommand{\ei}{\end{itemize}}

\newcommand{\ben}{\begin{enumerate}}
\newcommand{\een}{\end{enumerate}}
\newcommand{\ol}{\overline}

\def\){\right)}
\def\({\left( }
\def\]{\right] }
\def\[{\left[ }
\newcommand{\la}{\langle}
\newcommand{\ra}{\rangle}

\def\NO{\nonumber}

\newcommand{\be}{\begin{equation}}
\newcommand{\ee}{\end{equation}}

\def\bea{\begin{eqnarray}}
\def\eea{\end{eqnarray}}

\def\bal#1\eal{\begin{align}#1\end{align}}

\def\bald{\begin{aligned}}
\def\eald{\end{aligned}}

\def\beqx{\begin{displaymath}}
\def\eeqx{\end{displaymath}}

\newcommand{\bmat}{\left(\begin{array}}
\newcommand{\emat}{\end{array}\right)}

\def\d{\delta}
\def\e{\epsilon}
\def\f{\phi}
\def\g{\gamma}
\def\h{\eta}

\def\j{\psi}
\def\k{\kappa}

\def\m{\mu}
\def\n{\nu}

\def\p{\pi}

\def\s{\sigma}

\def\x{\xi}
\def\z{\zeta}

\def\F{\Phi}
\def\G{\Gamma}
\def\J{\Psi}

\def\vf{\varphi}

\def\cb{{\cal B}}

\def\cf{{\cal F}}
\def\cg{{\cal G}}

\def\cm{{\cal M}}
\def\cn{{\cal N}}
\def\co{{\cal O}}

\def\cv{{\cal V}}
\def\cw{{\cal W}}

\def\bo{{\raise-.3ex\hbox{\large$\Box$}}}               % D'Alembertian
\def\pa{\partial}                                       % curly d
\def\face{{\raise.2ex\hbox{$\displaystyle \bigodot$}\mskip-2.2mu \llap {$\ddot
        \smile$}}}                                   % happy face
\def\>{\rangle}                                      %right angle
\def\<{\langle}                                      %left angle

\def\wt#1{\widetilde{#1}}                            % big tilde
\def\lbar#1{\ensuremath{\overline{#1}}}              % big bar
\def\leftrightarrowfill{$\mathsurround=0pt \mathord\leftarrow \mkern-6mu
        \cleaders\hbox{$\mkern-2mu \mathord- \mkern-2mu$}\hfill
        \mkern-6mu \mathord\rightarrow$}        % <--> double differential
\def\dvec#1{\vbox{\ialign{##\crcr
        \leftrightarrowfill\crcr\noalign{\kern-1pt\nointerlineskip}
        $\hfil\displaystyle{#1}\hfil$\crcr}}}           % <--> accent
\def\-{\hphantom{-}}

\catcode`@=12

\numberwithin{equation}{section}

\begin{document}
\begin{titlepage}

\begin{flushright}
SISSA 40/2015/FISI
%ULB-TH/09-10\\
%hep-th/yymmnnn\\
\end{flushright}
\bigskip
\def\thefootnote{\fnsymbol{footnote}}

\begin{center}
\vskip -10pt
{\LARGE
{\bf
A goldstino at the bottom of the cascade%On supersymmetry  breaking vacua\\
%\vspace{0.30in}
% in cascading gauge theories
}
}
\end{center}

\bigskip
\begin{center}
{\large %Authors
Matteo Bertolini$^{1,2}$, Daniele Musso$^2$, \\ \vskip 5pt Ioannis Papadimitriou$^1$,  
Himanshu Raj$^1$}

\end{center}

\renewcommand{\thefootnote}{\arabic{footnote}}

\begin{center}
\vspace{0.2cm}
$^1$ {SISSA and INFN - Sezione di Trieste\\
Via Bonomea 265; I 34136 Trieste, Italy\\}
$^2$ {Abdus Salam International Centre for Theoretical Physics (ICTP)\\
Strada Costiera 11; I 34014 Trieste, Italy}

\vskip 5pt
%{\texttt{bertmat,dipietro,fporri @sissa.it}}

\end{center}

\noindent
\begin{center} {\bf Abstract} \end{center}
\noindent
Working within a five-dimensional consistent truncation of type IIB supergravity dimensionally reduced on $T^{1,1}$, we consider supersymmetry breaking solutions with the asymptotics of the supersymmetric KS background \cite{Klebanov:2000hb}. There exists a two-parameter family of such solutions. Within this family, we show that those (and only those) solutions related to antiD-branes at the tip of the conifold correspond to dual field theory vacua where a goldstino mode is present and supercurrent Ward identities hold. Our findings do not depend on the IR singularity of the dual backgrounds, nor on its resolution. As such, they constitute an independent, necessary check for the existence of supersymmetry breaking vacua in the conifold cascading gauge theory. Our analysis relies on a holographic derivation of the Ward identities which has a wider applicability, beyond the specific system and symmetries considered here.

\vspace{1.6 cm}
\vfill

\end{titlepage}

\setcounter{footnote}{0}

%\tableofcontents

%%%%%%%%%%%%%%%%%%%%%%%%%%%%%%%%%%%%%%%%%%%%%
%%%%%%%%%%%%%%%%%%%%%%%%%%%%%%%%%%%%%%%%%%%%%%
\section{Introduction and summary of the results}\label{sec1}

Since the early days of the AdS/CFT correspondence \cite{Maldacena:1997re,Witten:1998qj,Gubser:1998bc}, the 
new tools that have become available to understand field theory dynamics in the strong coupling regime 
have opened-up new promising avenues to study supersymmetric theories 
where supersymmetry is broken dynamically.

There is by now rather strong evidence that a large class of supersymmetric field theories admitting supersymmetry-breaking
vacua can be constructed in string theory.
We are thinking in particular of quiver gauge theories obtained by placing stacks of D-branes at Calabi-Yau singularities. 
This can be interesting in view of phenomenological applications within string compactification scenarios, but can also 
be instrumental within the gauge/gravity duality. Indeed, in the decoupling limit, one can have a way to describe, at least 
in principle, strongly coupled supersymmetry breaking vacua 
by means of dual gravitational backgrounds. This is promising, but in general more work is needed to have precise control on these vacua, understand their stability properties, dynamics and spectrum.

A concrete proposal to construct supersymmetry-breaking vacua in string theory
was put forward time ago in \cite{Kachru:2002gs} (from now on KPV)
for the ${\cal N}=1$ theory obtained by placing $N$ regular and $M$ fractional D3-branes at a conifold singularity. 
This is a quiver gauge theory with $SU(N+M) \times SU(N)$ gauge group, four bi-fundamental fields $A_i, B_j$ ($i,j=1,2$) 
and a quartic superpotential $W = \lambda \,\e^{ij} \e^{kl}\, \mbox{Tr} \, (A_i B_k A_j B_l)$ 
\cite{Klebanov:1999rd,Klebanov:2000nc,Klebanov:2000hb} (henceforth KS model). 
The proposal, based on the idea of adding 
antiD-branes at the tip of the deformed conifold, suggests that besides supersymmetric vacua, like the one described by the 
KS solution \cite{Klebanov:2000hb}, the dual field theory admits also supersymmetry-breaking, metastable vacua.  
If correct, this is likely not to be a specific phenomenon of the KS model, but rather a generic fact 
in D-brane/string constructions, see for instance \cite{Argurio:2006ny,Argurio:2007qk}. 
As a consequence, an understanding of the non-supersymmetric dynamics of the conifold theory 
has a more general relevance and it is not just interesting \textit{per se}.

In the gauge/gravity duality framework, a vacuum of the QFT is described by a 
(four-dimensional Poincar\'e invariant) five-dimensional solution of the dual gravitational system.
Solutions sharing the same asymptotics correspond, in general, to different vacua of the same QFT. 
A supergravity solution describing, asymptotically, the KPV vacuum was obtained  in \cite{DeWolfe:2008zy}. 
This solution, as the original one found in \cite{Klebanov:2000hb}, asymptotes to the Klebanov-Tseytlin (KT) solution \cite{Klebanov:2000nc} near the boundary. The latter, 
in fact, furnishes a UV-regulator for any gravitational background describing a vacuum of the KS theory.

In a QFT, whenever a global symmetry is spontaneously broken, a massless particle appears in the spectrum. 
In the case of supersymmetry, this is a fermionic mode, the goldstino. 
Hence, a natural question to try to answer is whether the supergravity mode dual to the goldstino 
field is present in the non-supersymmetric background of \cite{DeWolfe:2008zy}. 

When one deals with a supergravity solution which breaks supersymmetry, 
two obvious questions arise:
\begin{enumerate}
\item Is  the solution (meta)stable, gravitationally? 
\item Is the supergravity mode dual to the goldstino present? 
\end{enumerate}
A positive answer to the first question guarantees that the solution is describing holographically an actual QFT vacuum. 
The second ensures that in such a vacuum supersymmetry is broken spontaneously

From a QFT perspective, it is obvious that these two questions can be answered independently. 
The goldstino is the lowest energy excitation in the supercurrent operator $S_{\m\alpha}$,
and as such it appears as a massless pole in the two-point function
\begin{equation}
\label{sbars}
\langle S_{\m\alpha} \, \bar S_{\nu\dot\beta} \rangle~.
\end{equation}
This correlator has in general a very complicated structure, which depends on the vacuum that one is considering. 
However, in order to display the goldstino pole, one does not need to compute \eqref{sbars} fully. 
The information is encoded just in the term implied by the supersymmetry Ward identity
\begin{align}
\label{wiS}
\langle \partial^\m S_{\m\alpha}(x) \,\bar S_{\nu\dot\beta}(0) \rangle =  - 2  \sigma^\m_{\ \alpha\dot\beta} \, \langle T_{\m\nu}\rangle \, \d^4(x) ~,
\end{align}
which is a (quasi-local) contact term. (Upon integration, this identity relates the vacuum energy $E \sim \eta^{\m\nu} T_{\m\nu}$ 
to the residue of the goldstino pole in the two-point function \eqref{sbars} \cite{Argurio:2013uba}.)
Ward identities hold in any vacuum of a  QFT, and depend on UV data only. On the contrary, vacuum stability is an IR property. 

For theories with a gravity dual, this disentanglement should emerge from a holographic analysis, too. 
In  \cite{Argurio:2014uca} a rather general class of holographic supersymmetric RG-flows was considered, 
Ward identities as \eqref{wiS} were derived holographically, and it was shown that, indeed, they hold 
regardless of the detailed structure of the bulk solution in the deep interior, the presence of IR singularities and their possible resolution mechanism.\footnote{Similar results were obtained for bosonic global symmetries in \cite{Argurio:2014rja}.} 

Whenever one has sufficient control on the QFT, this result can be seen (just) as a consistency check of the AdS/CFT correspondence.
But it may become instrumental when one has to deal with field theories for which a satisfactory understanding of the dynamics and vacuum structure is lacking.
The KS theory falls in this class, at least as far as supersymmetry-breaking vacua are concerned. 
There has been a lively discussion in the last few years, initiated in \cite{Bena:2009xk}, 
regarding the stability properties of the dual supersymmetry-breaking backgrounds and the mechanism 
to resolve the IR singularity.\footnote{See the citation list for  \cite{Bena:2009xk} for a complete account of the many contributions since then. Suggestive results in favor of (meta)stability of the KPV vacuum were recently obtained in \cite{Michel:2014lva} working within an effective field theory approach. For a discussion regarding the possibility to cloak the singularity beyond an event horizon, instead, which  according to the criterion of \cite{Gubser:2000nd} would make it acceptable, see \cite{Hartnett:2015oda,Cohen-Maldonado:2015ssa}.} In this work we do not offer any new insight on this issue. 
What we do, instead, will be to apply the  analysis of \cite{Argurio:2014uca} to the KS model, and try to 
give a definite answer to the second question. The answer will be affirmative.
In particular, we derive via holography the supercurrent Ward identities \eqref{wiS} for the KS cascading theory, 
and, by computing explicitly eq.~\eqref{wiS}  both in supersymmetric and supersymmetry-breaking vacua, 
we find the goldstino pole whenever expected. 
Our results confirm the possibility that spontaneous supersymmetry-breaking vacua may exist in the KS model, specifically that 
a goldstino mode is indeed present in the asymptotic solution of \cite{DeWolfe:2008zy}. 
As an interesting outcome of our analysis, we show that some recently-found non-supersymmetric supergravity solutions 
\cite{Kuperstein:2014zda}, which have an asymptotic compatible with the KS theory, do not accommodate a goldstino mode. Hence they correspond to explicit, rather than spontaneous, supersymmetry breaking.

Holographic renormalization for cascading theories is known to be trickier than for asymptotically AdS (AAdS) backgrounds, 
and we will clarify a couple of issues which are instrumental to holographically renormalize the theory in these cases. 
In particular, we will argue that to treat the log-divergent structure of cascading backgrounds properly, it is appropriate 
to define the renormalized action in terms of induced fields instead of the sources, define the renormalized correlators as functions of 
induced fields at the cut-off \cite{Papadimitriou:2004ap}, and take the cut-off to infinity only at the very end of the calculation \cite{Freedman:1998tz}. 

The rest of the paper is organized as follows. We start in section 2 by presenting the relevant five-dimensional supergravity Lagrangian, and derive supersymmetric and non-supersymmetric solutions with correct KS asymptotics. 
The latter are a two-parameter family. Although these are known results, we re-derive them from a consistently truncated 5d supergravity, 
as a preliminary step for the subsequent analysis. In section 3, which contains the main results of our paper, we derive holographically all 
the supersymmetry Ward identities that we need, showing that they hold independently of the vacuum one considers. 
The derivation, which relies on the existence of local covariant counterterms that renormalize the on-shell action, as well as on a renormalization scheme respecting boundary diffeomorphisms and supersymmetry transformations, 
is general enough to account for any Ward and operator identities one expects to hold.\footnote{The renormalization scheme, though, will generically break Weyl and superWeyl invariance, leading to a trace and a supertrace anomaly, respectively.} 
Finally, in section 4 we evaluate explicitly the supersymmetry Ward identities for those vacua described by the solutions derived in section 2. 
The requirement of non-vanishing vacuum energy selects only a one-dimensional subspace within the space of supersymmetry-breaking solutions, 
in agreement with the analysis of \cite{DeWolfe:2008zy}, where evidence was also given that this corresponds to the set of (asymptotic) 
solutions generated by antiD-branes at the tip of the conifold. The field/operator map will offer a simple explanation of these results 
from the dual field theory perspective, including the absence of a goldstino mode for the complementary set of solutions. 
In section 5 we present our conclusions and outlook. Several appendices contain a number of technical details that we omitted from the main body of the paper.

%%%%%%%%%%%%%%%%%%%%%%%%%%%%%%%%%%%%%%%%%%%%%%
%%%%%%%%%%%%%%%%%%%%%%%%%%%%%%%%%%%%%%%%%%%%%%
\section{Cascading theories from 5d supergravity}\label{sec2}

The 5d ${\cal N}=2$ supergravity that we need is obtained by reducing 10d type IIB supergravity on $T^{1,1}$, the conifold basis. 
The supergravity theory that one should consider in order to analyze the full KS cascading theory 
(namely, to describe its complete set of vacua) is rather complicated; 
almost intractable, in fact. However, there are a number of simplifications that our analysis allows. 

First, we will focus on an $SU(2) \times SU(2)$-invariant truncation (the dimensional reduction was performed in \cite{Cassani:2010na} and \cite{Bena:2010pr}; we use the notations of \cite{Cassani:2010na}). This truncation cannot capture all possible vacua of the KS theory, but it is general enough to admit the original KS solution 
as one of its supersymmetric solutions. This solution describes the most symmetric point in 
the baryonic branch of the $SU(N+M) \times SU(N)$ KS model, with $N=kM$ and $k$ an integer number. 
The same bulk Lagrangian admits also supersymmetry-breaking solutions, some of which should describe, according 
to the KPV construction, a metastable vacuum of the $SU(N+M) \times SU(N)$ cascading theory with vacuum energy 
$E \sim p$, where now $N = kM -p$, and $p \ll M$ (in the KPV vacuum $p$ corresponds to the number of antiD-branes; 
from a ten-dimensional viewpoint, keeping the $SU(2)\times SU(2)$ symmetry amounts to smearing the antiD-branes 
over the compact space). In fact, we will work with a simplified ansatz, which preserves an extra $U(1)$ symmetry 
\cite{Buchel:2014hja} and which can just accommodate KT-like solutions. 
This simplification further reduces the number of active fields
and, in particular, it excludes the mode related to the conifold deformation parameter.

A second simplification occurs at the level of the solutions themselves. 
As already stressed, in order to prove the presence of the goldstino, one does not need to consider the full solution 
but just its asymptotic expansion up to the order where the supersymmetry-breaking deformation appears. 
This simplifies the analysis considerably, and allows one to consider the backgrounds only to order $z^4$, $z$ being the holographic coordinate. 
This may sound inconsistent, at first sight. Indeed, as noticed in \cite{DeWolfe:2008zy,Bena:2011wh}, 
the KS and KT solutions, which are one and the same to leading order in a near-boundary expansion, 
differ already at order $z^3$ by terms proportional to $\varepsilon$, the conifold deformation parameter, which is zero in the KT solution.
These effects are dominant against $z^4$, the order at which supersymmetry-breaking effects enter. 
However, being a supersymmetric deformation, it is possible to see that $\varepsilon$ does not affect
the supersymmetry-breaking dynamics in any dramatic 
manner, modifying, at most, the numerical values of some quantities, but not the possible existence 
of supersymmetry-breaking vacua and of the associated massless fermionic mode. 

Table \ref{massSpectrum} contains all the fields entering the truncation and the multiplet structure, including, for future reference, 
the AdS masses obtained in the conformal limit \cite{Klebanov:1998hh}, $M=0$. We refer to appendix \ref{sigma-model} for more details on the five-dimensional $\sigma$-model.

\begin{table}[h]
\begin{center}
$	\begin{array}{lcccccl}\hline
\mathcal N=2\: {\rm multiplet} && {\rm field\: fluctuations} && \hskip-0.3cm\tx{AdS mass}   \\\hline
\rule{0pt}{3ex}
{\rm gravity} && \begin{array}{c}V_A\\ \J_{A} \\ g_{AB} \end{array} && \hskip-0.3cm\begin{array}{l} m^2=0 \\m={3\over 2} \\ m^2=0 \end{array}  \\ \hline
{\rm universal \;hyper} && \begin{array}{c} b^\Omega - i \,c^\Omega \\ \z_\f\\ \tau=C_0+ie^{-\f} \end{array} && \begin{array}{l} m^2=-3 \\m=-{3\over 2}\\ m^2=0 \end{array}  \\ \hline
{\rm Betti \;hyper} &&	\begin{array}{c}	 t \,e^{i\theta}\\ \z_b\\ b^\F,\;\;c^\F \end{array} && \begin{array}{l}m^2= -3 \\m=  -{3\over 2}\\m^2= 0   \end{array}   \\  \hline
{\rm massive \;vector} && \begin{array}{c} V\\ \z_V\\V_A^1\\ b^\Omega + i \,c^\Omega \\ \z_U\\ U \end{array} && \begin{array}{l} m^2=12\\m={9\over2} \\m^2= 24 \\m^2= 21\\m=-{11\over2}\\m^2= 32  \end{array}   \\ \hline
\end{array}$
\caption{Spectrum of bosons and fermions in the ${\cal N}=2$ truncation of \cite{Cassani:2010na} (5d indices  are dubbed $A,B$).}
\label{massSpectrum}
\end{center}
\end{table}

To search for domain wall solutions, we can truncate the Lagrangian to its scalar field content only (plus the graviton). Moreover, the extra $U(1)$ symmetry reduces the number of active scalar fields to just four, which, without loss of generality, we can take to be real. The end result is
\begin{equation}\label{sugra}
S=  \int d^5x\sqrt{-g_5}\,\(R-\frac{8}{15}dU^2-\frac{4}{5} dV^2- e^{-{4\over 5}(U+V)-\f}\(db^\F\)^2-{1\over 2} d\f^2-{\cal V}\) \ ,
\end{equation}
where we have set the five-dimensional Newton constant $G_5 = 1/16 \pi$ 
and $R$ is the Ricci scalar. The scalar potential $\cal V$ is given by
\be
\label{pot}
{\cal V}={1\over 2} (27\pi N - 9M~b^\F)^2e^{-{8\over 3}U}+ {81\over 4}M^2e^{-{4\over 15}(7U-3V)+\f}-24 e^{-{2\over 15}(8U+3V)}+4 e^{-{4\over 15}(4U+9V)}\ .
\ee
The parameters $N$ and $M$ are continuous quantities in supergravity, but
should be thought of as integers, since they correspond to type 
IIB higher-form fluxes integrated over the non-trivial cycles of $T^{1,1}$ and are thus quantized.
Upon uplifting, they are related respectively to the number of regular and fractional D3-branes at the conifold singularity.

%%%%%%%%%%%%%%%%%%%%%%%%%%%%%%%%%%%%%%%%%%%%%%
%\newpage
\subsection{Supersymmetric and non-supersymmetric solutions}
\label{susynonsusy}

The solutions we are after should correspond to vacua of the KS dual field theory and, as such, 
they should satisfy given boundary conditions.
First, due to four-dimensional Poincar\'e invariance, we should focus on domain wall solutions,
where all scalars depend on the radial coordinate only and where the ansatz for the metric reads
\be
\label{ansa}
ds^2=\frac 1{z^2} \left(e^{2Y(z)}\eta_{\m\nu}dx^\m dx^\nu +e^{2X(z)}dz^2\right)\ ,
\ee
with $\m,\nu=0,\dots 3$. The function $X(z)$ can be eliminated by a redefinition of the radial coordinate, while the function $Y(z)$ is the only dynamical variable parameterizing the domain wall metric. 
From now on we split the 5d indices as $A=(z,\m)$.
The AdS metric is recovered for $X=Y=0$, the conformal boundary being at $z=0$. 
Another requirement is that for $M=0$ we should recover the Klebanov-Witten (KW) AdS solution \cite{Klebanov:1998hh}.

The solutions we derive below were already obtained working in a ten-dimensional setting in 
\cite{DeWolfe:2008zy} (see also \cite{Aharony:2005zr} whose normalization for the metric is the same as ours). 
In this section we re-obtain the same solutions within the truncated five-dimensional model \eqref{sugra}. 

Imposing that the fields satisfy the BPS equations (see appendix \ref{sigma-model}) one finds the supersymmetric solution
\begin{align}
e^{2Y}& =h^{1\over 3}(z) \ , \
e^{2X}=h^{4\over 3}(z) \ , \
e^{2U}=h^{5\over 2}(z)\ , \nn \\
b^\F(z)&=-{9\over 2}g_sM~\log \(z/z_{0}\)\ , \nn \\
\f(z)&= \tx{log}~g_s \ , \ V=0 \ ,
\label{KTsol}
\end{align}
where the warp factor $h(z)$ is
\be
\label{hbps}
h(z)=\frac{27\pi}{4g_s}\(g_sN+{1\over 4}a (g_sM)^2 -a (g_sM)^2 \log \(z/z_{0}\)\), \,
\ee
with $a=3 / 2\pi$, and $z_{0}$ is a scale introduced to make the arguments of the log's dimensionless
(in the dual QFT, $z_{0}$ corresponds to a renormalization scale). 
The parameter $g_s$, which in 5d supergravity is an integration constant, has been dubbed as the 10d string coupling, 
to which it actually gets matched upon uplifting. The characteristic features of this solution are a constant dilaton $\f$ and a vanishing 
$V$ field.
This solution is nothing but the five-dimensional formulation of the KT-solution \cite{Klebanov:2000nc}. 
The KW pure AdS solution \cite{Klebanov:1998hh} is recovered upon setting $M=0$. 

We now look for solutions of the second order equations of motion descending from the action \eqref{sugra} 
(see again appendix \ref{sigma-model}). We should require that the solutions reduce to the supersymmetric solution \eqref{KTsol}-\eqref{hbps}  
in the far UV, that is as $z \rightarrow 0$. Up to the order $z^4$, which is our focus here, 
the general solution depends on two additional parameters only which, adapting to the notation of \cite{DeWolfe:2008zy}, 
we denote with ${\cal S}$ and $\varphi$.
The result is 
\begin{align}
e^{2Y}&= h^{1\over 3}(z)\,h_2^{1\over 2}(z)\,h_3^{1\over 2}(z) \quad , \quad 
e^{2X} = h^{4\over 3}(z)\,h_2^{1\over 2}(z) \ ,\nn \\
e^{2U}&= h^{5\over 2}(z)\,h_2^{3\over 2}(z) \quad , \quad
e^{2V} = h_2^{-\frac{3}{2}}(z)\ , \nn \\
b^\F(z)&= -{9\over 2}g_sM~\log \(z/z_{0}\) \nn \\
&+ z^4 \[\({9\pi N \over 4M}+{99\over 32}g_sM-{27\over 4}g_sM\log \(z/z_{0}\)\){\cal S}-{9\over 8} g_sM\varphi\]+{\cal O}(z^8)\ ,  \nn \\
\f(z)&= \log g_s+z^4\(3 {\cal S}\log \(z/z_{0}\)+\varphi\)+{\cal O}(z^8) ~,
\label{nonSUSYbg}
\end{align}
where
\begin{align}
&h(z)=\frac{27\pi}{4g_s}\(g_sN+{1\over 4}a (g_sM)^2 -a (g_sM)^2 \log \(z/z_{0}\)\) \\
&+\frac{z^4}{g_s}\[\({54\pi g_s N \over 64}+{81\over4}{13\over 64}\(g_sM\)^2-{81\over 16}\(g_sM\)^2 \log \(z/z_{0}\)\){\cal S}-{81\over 64}\(g_sM\)^2 \varphi\]+{\cal O}(z^8)\ ,\nn \\
&h_2(z)= 1+\frac{2 }{3}{\cal S} z^4+{\cal O}(z^8)\quad , \quad h_3(z)=1+{\cal O}(z^8) ~.
\end{align}
This two-parameter family breaks supersymmetry, in general, but reduces to 
the supersymmetric KT solution of \eqref{KTsol}-\eqref{hbps} for ${\cal S}=\varphi=0$.
Furthermore, as anticipated, supersymmetry-breaking effects enter at order $z^4$ relative to the KT solution, 
so for $z\rightarrow 0$ the generic solution within the two-parameter family asymptotes to KT. 
Note, moreover, that the dilaton now runs.  
In \cite{DeWolfe:2008zy} evidence was given that the branch $\varphi=0$ describes (the large distance asymptotics of) 
the solution generated by $p$ antiD3-branes at the tip of the conifold, ${\cal S}$ being proportional to $p$. 
On the contrary, the  branch ${\cal S}=0$, which in the AdS limit $M=0$ corresponds to the usual independent fluctuation 
of the dilaton \cite{Gubser:1999pk,Kehagias:1999tr}, was recently extended to all orders in $z$ and a full (still singular) 
solution was found \cite{Kuperstein:2014zda}.\footnote{The matching between the branch ${\cal 
S}=0$ and the solution of \cite{Kuperstein:2014zda} can be seen upon the following relation between the parameters 
$\varphi=-\sqrt{10}\,r_s^4$, while the holographic coordinates are inverse to one another, $z=1/r$.} 
As we will see later, this branch describes a vacuum where supersymmetry is explicitly broken in the dual 
field theory and hence does not correspond to a vacuum of the KS field theory. Let us finally notice, in passing, 
that an ansatz with constant $h_3(z)$ is inconsistent with the equations of motion. 
Although $h_3(z)$ does not affect the solution at order $z^4$, one can check that it is necessary to have $h_3(z)$
non-trivial at order $z^8$ in order to extend the solution deeper in the bulk.

%%%%%%%%%%%%%%%%%%%%%%%%%%%%%%%%%%%
%%%%%%%%%%%%%%%%%%%%%%%%%%%%%%%%%%%
\section{Holographic Ward identities}
\label{HWI}
The KS theory is an ${\cal N}=1$ QFT and supersymmetry Ward identities like \eqref{wiS} should hold in any of its vacua. 
In this section we provide a holographic derivation of these identities. In the next section, we will test them against 
the supersymmetric and non-supersymmetric solutions that we have found in section \ref{susynonsusy}. 

Fields in the bulk are dual to QFT gauge invariant operators. In the present case, the bosonic bulk sector consists of four real scalars and the metric. 
In particular, the fields $e^{-\f}$ and $\wt b^\F = e^{-\f}\, b^\F$ are dual to dimension 4 operators, ${\cal O}_\f$ and ${\cal O}_{\wt b}$  (see e.g. \cite{Klebanov:1999rd}), which are respectively related to the sum and the difference of the inverse of the two gauge couplings squared.\footnote{In fact, the precise correspondence involves also the quartic superpotential coupling \cite{Klebanov:1998hh,Klebanov:2000hb}.} In the conformal limit they are exact moduli. The scalars $V$ and $\wt U=(4qb^\F-4k+q^2e^\f)e^{-\frac45U}/8$ are dual to dimension 6 and 8 operators, respectively. The constants $k$ and $q$ are defined in \eqref{conversion}. As explained in appendix \ref{hol-map}, the composite field $\wt U$ is the unique combination of bosonic fields that is sourced solely by the dimension 8 operator. 
Moreover, although not necessary, it is natural to define the covariant source of the energy-momentum tensor as the field that couples only to metric fluctuations, namely $\wt\g_{\m\n}=e^{-4U/15}\g_{\m\n}$, where $\g_{\m\n}$ is the four-dimensional induced metric at the radial cut-off. The fermionic sector contains four spin $1/2$ fermions and the spin $3/2$ gravitino. The field $\wt\J_\m^+=e^{-\frac{2}{15}U}\left(\J_\m^+-\frac{2i}{15}\G_\m\z_U^-\right)$ is dual to the supercurrent, the supersymmetric partner of the energy-momentum tensor, while the fields $\z_\f$ and $\wt \z_b = e^{-\f}\left(\z_b-b^\F\z_\f\right)$, are dual to dimension $7/2$ operators, the supersymmetric partners of ${\cal O}_\f$ and ${\cal O}_{\wt b}$, respectively.  Finally, $\z_V$ and $\wt\z_U=-\frac 45 \wt U\z_U+\frac18e^{-\frac45 U}\left(4q\z_b+q^2e^\f\z_\f\right)$ are dual to irrelevant operators as their 
supersymmetric partners $V$ and $\wt U$. More details on the identification of the bulk fields dual to gauge-invariant operators can be found in appendix \ref{hol-map}. In what follows, we will switch off the sources of bulk fields that are dual to irrelevant operators. 
Moreover, the asymptotic supersymmetry breaking solution we presented in section \ref{susynonsusy} is given just to order $z^4$, and this is sufficient for calculating VEVs of relevant 
or marginal operators only.\footnote{We could turn on a (perturbative) source for the irrelevant operators and calculate their VEVs once 
we obtain an asymptotic solution to order $z^8$.}

As a first step towards the derivation of the Ward identities, we have to define holographically the renormalized one-point functions in the presence of sources. 
The former are defined as derivatives of the renormalized on-shell action at a radial cut-off with respect to the induced fields at the cut-off and read  
(care is required here since, as we have already noticed, the supergravity field basis is not diagonal with respect to the basis of the field theory operators)
\begin{align}
\label{oneptf}
&\langle T^{\m\n} \rangle =\frac{2}{\sqrt{-\wt\g}}\left.\frac{\del S_\tx{ren}}{\del \wt\g_{\m\n}}\right|_{\f,\wt b^\F,\wt U,\wt\J^+,\z_\f^-,\wt\z_b^-,\wt\z_U^-},
% % % % % % % % % % % % % % % % % % % % % % % % % % % % % % % % % % %
&&\langle \ol{S}^{-\m} \rangle =\frac{-2i}{\sqrt{-\wt\g}}\left.\frac{\del S_\tx{ren}}{\del \wt\J_\m^+}\right|_{\wt\g,\f,\wt b^\F,\wt U,\z_\f^-,\wt\z_b^-,\wt\z_U^-},\NO\\ %\NO\\
% % % % % % % % % % % % % % % % % % % % % % % % % % % % % % % % % % % %
&\langle {\cal O}_\f \rangle =\frac{1}{2\sqrt{-\wt\g}}\left.\frac{\del S_\tx{ren}}{\del \f}\right|_{\wt\g,\wt b^\F,\wt U,\wt\J^+,\z_\f^-,\wt\z_b^-,\wt\z_U^-}, %\NO\\
% % % % % % % % % % % % % % % % % % % % % % % % % % % % % % % % % % % 
&&\langle \ol{\cal O}^{+}_{\z_\f} \rangle =\frac{1}{\sqrt{-\wt\g}}{i\over \sqrt{2}}\left.\frac{\del S_\tx{ren}}{\del \z_\f^-}\right|_{\wt\g,\f,\wt b^\F,\wt U,\wt\J^+,\wt\z_b^-,\wt\z_U^-},\NO\\ %\NO\\
% % % % % % % % % % % % % % % % % % % % % % % % % % % % % % % % % % % %
&\langle {\cal O}_{\wt b} \rangle =\frac{1}{2\sqrt{-\wt\g}}\left.\frac{\del S_\tx{ren}}{\del \wt b^\F}\right|_{\wt\g,\f,\wt U,\wt\J^+,\z_\f^-,\wt\z_b^-,\wt\z_U^-},  %\nn \\
% % % % % % % % % % % % % % % % % % % % % % % % % % % % % % % % % % % 
&&\langle \ol{\cal O}^{+}_{\wt\z_b} \rangle =\frac{1}{\sqrt{-\wt\g}}{i\over \sqrt{2}}\left.\frac{\del S_\tx{ren}}{\del \wt\z_b^-}\right|_{\wt\g,\f,\wt b^\F,\wt U,\wt\J^+,\z_\f^-,\wt\z_U^-},
\end{align}
where the subscripts in the partial functional derivatives indicate the variables held fixed, which is crucial for evaluating correctly these one-point functions. The resulting expressions in terms of derivatives with respect to the supergravity fields are given in appendix \ref{hol-map}. The quantity $\wt\g$ is the determinant of $\wt\g_{\m\n}$, while the normalization of the one-point functions has been chosen in accordance with the conventions for organizing these operators in $\cn=1$ superfields.

Several comments are in order here. Firstly, $S_\tx{ren}$ denotes the renormalized on-shell action
\be\label{Sren}
S_\tx{ren}=S_\tx{reg}+S_\tx{ct}\ ,
\ee 
where the regularized action $S_\tx{reg}$ stands for the bulk on-shell action plus the Gibbons-Hawking term (together with its supersymmetric completion \cite{Argurio:2014uca}), and the covariant boundary counterterms $S_\tx{ct}$ contain both bosonic and fermionic terms. 
The counterterms, by construction, ensure that $S_\tx{ren}$ admits a smooth limit as the radial cut-off is removed. 
Given the asymptotic behavior of the induced fields given in appendix \ref{eoms}, this implies that the renormalized one-point functions with the cut-off removed correspond to the limits
\bal\label{QFT-1pt-fns}
\langle T^\m_\n \rangle_\tx{QFT} &=\lim_{z\to 0}z^{-4}\langle T^\m_\n \rangle\ ,\quad 
\langle \ol{S}^{-\m} \rangle_\tx{QFT} =\lim_{z\to 0}z^{-9/2}e^{-X(z)/8}\langle \ol{S}^{-\m} \rangle\ ,\NO\\
\langle {\cal O}_\f \rangle_\tx{QFT} &=\lim_{z\to 0}z^{-4}\langle {\cal O}_\f \rangle\ ,\quad \langle \ol{\cal O}^{+}_{\z_\f} \rangle_\tx{QFT} =\lim_{z\to 0}z^{-7/2}e^{-X(z)/8}\langle \ol{\cal O}^{+}_{\z_\f} \rangle\ ,\NO\\
\langle {\cal O}_{\wt b} \rangle_\tx{QFT} &=\lim_{z\to 0}z^{-4}\langle {\cal O}_{\wt b} \rangle\ ,\quad
\langle \ol{\cal O}^{+}_{\wt\z_b} \rangle_\tx{QFT} =\lim_{z\to 0}z^{-7/2}e^{- X(z)/8}\langle \ol{\cal O}^{+}_{\wt\z_b} \rangle\ .
\eal  
Note that one of the indices of the stress tensor has been lowered with the field theory metric $\wt\g_{\m\n}$, and not $\g_{\m\n}$. 
The explicit expression for the local boundary counterterms is not required in order to derive the Ward identities holographically. It suffices that there exist local and covariant boundary counterterms that render the on-shell action finite, while preserving the symmetries of the dual QFT --most importantly for us, supersymmetry-- up to possible anomalies. Of course, explicit knowledge of the counterterms is necessary in order to evaluate the one-point functions \eqref{QFT-1pt-fns} for any given solution. In the next section we will present the boundary counterterms required to evaluate the bosonic VEVs in domain wall backgrounds of the form \eqref{nonSUSYbg}, which is all we need for what we do in this paper. A systematic derivation of both bosonic and fermionic counterterms for generic cascading theories will be presented elsewhere \cite{usfut}.   

Another point worth mentioning is that the one-point functions of the bosonic operators are given by the derivative of the renormalized action with respect to the corresponding induced field on the radial cut-off, which is therefore  identified with the covariant source. However, the covariant sources for the fermionic operators are given by the corresponding induced field --which is a four-dimensional spinor-- projected onto a definite chirality. As a consequence, the dual operators have definite (and opposite) chirality.\footnote{This difference reflects the structure of the radial Hamiltonian phase space for bosonic and fermionic fields. The holographic one-point functions \eqref{oneptf} are in either case the renormalized radial canonical momenta \cite{Papadimitriou:2004ap}.} The chirality that corresponds to the covariant fermionic source is determined by the leading asymptotics which in turn are fixed by the sign of the their masses (see Table \ref{massSpectrum} and 
appendix \ref{eoms} for details). 
 
Given the holographic identification of the covariant sources and one-point functions at the radial cut-off, 
the derivation of the Ward identities proceeds exactly as in standard QFT textbooks. Namely, global symmetries 
are gauged, giving rise to generic sources for all global symmetry currents. In addition, sources are manually 
turned on for all other operators, such as scalar and fermion operators. 
Using the transformation of all the sources under the local (gauged) symmetries together with the invariance (up to anomalies) of the generating functional, 
leads to the Ward identities at the level of one-point functions in the presence of arbitrary sources. 
In the bulk description all symmetries are already gauged and all sources are turned on, so the only other 
ingredient we need in order to derive holographically the Ward identities is the transformation of the covariant 
sources under the local symmetries. These are given explicitly in appendix \ref{local-symmetries}. 
In the bulk these symmetries correspond 
to infinitesimal local supersymmetry transformations and bulk diffeomorphisms generated respectively by 
a 4-component Dirac spinor $\e$ and a 5-vector $\xi^A$, preserving the gauge-fixing conditions \eqref{gf-conditions}. 
The spinor $\e$ has 8 real components which correspond to the 8 real supercharges of the ${\cal N}=2$ 5d supergravity. 
This can be written as $\e=\e^++ \e^-$. Since $\e^+$ and $\e^-$ are linearly independent supersymmetry 
transformation parameters, the renormalized on-shell action is not only invariant under $\e$ but also under $\e^+$ and $\e^-$ independently. 
The spinor $\e^+$ generates (local) boundary supersymmetry transformations, while $\e^-$ generates superWeyl transformations. 
Invariance under $\e^+$ and $\e^-$ leads 
respectively to the supersymmetry Ward identities and the operator identity involving the gamma-trace of the supercurrent. 
Similarly, the infinitesimal bulk diffeomorphisms $\x ^A$ 
preserving the gauge-fixing conditions \eqref{gf-conditions} are parameterized by two 
independent parameters, a scalar $\s(x)$ generating boundary Weyl transformations\footnote{The corresponding bulk 
diffeomorphisms are known as Penrose-Brown-Henneaux (PBH) diffeomorphisms and are discussed in detail in \cite{Imbimbo:1999bj}.} 
and an infinitesimal boundary diffeomorphism $\x_o^\m(x)$. Invariance under these leads respectively to the trace 
Ward identity and the Ward identity  involving the divergence of the stress tensor. If the theory has a trace anomaly, 
then supersymmetry implies that there will also be an anomaly in the operator identity involving the gamma-trace of the supercurrent.

%%%%%%%%%%%%%%%%%%%%%%%%%%%%%%%%
\subsection{Supersymmetry Ward identities}
\label{HWI1}

The supersymmetry Ward identities are obtained by requiring the invariance of the renormalized action under the local spinor $\e^+$, $\d_{\e^+}S_\tx{ren}=0$. However, to calculate $\d_{\e^+}S_\tx{ren}$, we need the 
transformation properties of the covariant sources under $\e^+$, 
which are given in appendix \ref{local-symmetries}, eq.~\eqref{sources-e+} . Using the one-point functions \eqref{oneptf} the variation of the renormalized action under $\e^+$ gives\footnote{Note that there are no contributions to the Ward identities from the irrelevant operators dual to $V$ and $\wt U$, as well as their fermionic superpartners, because their sources can be consistently set to zero.}
\begin{align}
\hskip-0.2cm\d_{\e^+}S_\tx{ren} &=\int d^4x\sqrt{-\wt\g}\(\frac{i}{2}\langle\ol{S}^{-\m}\rangle\d_{\e^+}\wt\J_\m^++{1\over 2}\langle T^{\m\n}\rangle\d_{\e^+}\wt\g_{\m\nu}+2\langle {\cal O}_\f\rangle\d_{\e^+}\f + 2\langle {\cal O}_{\wt b} \rangle\d_{\e^+} \wt b^\F \) \nn \\
&=\int d^4x\sqrt{-\wt\g}\(-\frac{i}{2}e^{-\frac{2}{15}U}\langle\del_\m\ol{S}^{-\m}\rangle-{1\over 2}\langle T^{\m\n}\rangle\ol{\wt\J}_\m^+\wt\G_\n+  i \langle {\cal O}_\f\rangle\ol{\z}_\f^{-} +  i\langle {\cal O}_{\wt b}\rangle\ol{\wt\z}_b^{-}\)\e^+ = 0~,
\end{align}
which implies the following identity between one-point functions at non-zero sources 
\be
\frac{i}{2}e^{-\frac{2}{15}U}\langle\del_\m\ol{S}^{-\m}\rangle=-{1\over 2}\langle T^{\m\n}\rangle\ol{\wt\J}_\m^+\wt\G_\n+  i \langle {\cal O}_\f\rangle\ol{\z}_\f^{-}+  i \langle {\cal O}_{\wt b}\rangle\ol{\wt \z}_b^{-}~,
\ee
where $\wt\G_\m=\wt e_\m^a\g_a=e^{-\frac{2}{15}U} e_\m^a\g_a$.
We can now differentiate this identity with respect to the various fermionic fields, i.e. the covariant sources, and then put all sources to zero 
to obtain \footnote{Notice that the two-point functions in \eqref{sWI1} 
(and the ensuing equations) are defined in terms of the one-point functions as: 
$\langle \del_\m \ol{S}^{-\m} S^{-\n}\rangle=-\frac{-2i}{\sqrt{-\wt\g}}\frac{\d}{i \d \ol{\wt\J}^+_\n}\langle\del_\m\ol{S}^{-\m}\rangle$. 
The extra factor of $i$ in the denominator is because of the Lorentzian signature and the overall minus sign is because the 
functional derivative is with respect to a Grassmann variable.}
\begin{align}
e^{-\frac{2}{15}U}\langle \del_\m \ol{S}^{-\m} (x) S_\n^{-}(0)\rangle &= 2i\, \wt\G_\m\langle T^\m_\n\rangle\, \d^4(x,0)\ ,\label{sWI1}\\
e^{-\frac{2}{15}U}\langle \del_\m \ol{S}^{-\m}(x) {\cal O}^+_{\z_\f}(0) \rangle &= - \sqrt{2}\, \langle {\cal O}_\f\rangle \,\d^4(x,0)\ ,\label{sWI2} \\
e^{-\frac{2}{15}U}\langle \del_\m \ol{S}^{-\m}(x) {\cal O}^+_{\wt\z_b}(0) \rangle &= - \sqrt{2}\, \langle {\cal O}_{\wt b}\rangle \,\d^4(x,0)\label{sWI3}~,
\end{align}
where $\d^4(x,y)=\d^4(x-y)/\sqrt{-\wt\g}$ is the covariant 4d Dirac delta function.
The last step is to take the cut-off all the way to infinity, which can be done using the limits \eqref{QFT-1pt-fns}.  All these limits can be easily evaluated using the asymptotic expansions of the induced fields given in appendix \ref{eoms}. 
Notice that all fermionic operators here are in the Dirac representation. In order to match with the field theory expressions, it 
is better to convert them into Weyl notation. This can be done easily using the following conversion rules 
\be
\psi^+=\psi_\alpha,~~~~\psi^-=\ol{\psi} ^{\dot{\alpha}}, ~~~~\ol{\psi}^+=\ol{\psi}_{\dot{\alpha}},~~~~\ol{\psi}^-=\psi^{\alpha}~,~~~~\(\g_\m\)_{\alpha\dot{\beta}}=i\(\sigma_\m\)_{\alpha\dot{\beta}}.
\ee
Adopting the above dictionary and upon sending the cut-off to infinity, we eventually get 
\begin{align}
\label{wiS1}
\la \del^\m S_{\m \alpha}(x)~\bar{S}_{\nu\dot{\beta}}(0)\ra_\tx{QFT} &=-2\, \sigma^\m_{\alpha\dot{\beta}}\la T_{\m\nu} \ra_\tx{QFT}~\d^4(x)\ ,\\
\label{wiS2}
\la \del^\m S_{\m}^\alpha(x)~{\cal O}_{\z_\f \alpha}(0)\ra_\tx{QFT} &= - \sqrt{2}~\la {\cal O}_\f \ra_\tx{QFT}~\d^4(x)\ , \\
\label{wiS3}
\la \del^\m S_{\m}^\alpha(x)~{\cal O}_{\wt \z_b \alpha}(0)\ra_\tx{QFT} &= - \sqrt{2}~\la {\cal O}_{\wt b} \ra_\tx{QFT}~\d^4(x)~.
\end{align}

The identity \eqref{wiS1} reproduces exactly the supercurrent Ward identity \eqref{wiS}. Eqs. \eqref{wiS2} and \eqref{wiS3} 
are analogous Ward identities for the supermultiplets where the operators ${\cal O}_\f$  and ${\cal O}_{\wt b}$ sit. 
Since ${\cal O}_\f$ and ${\cal O}_{\wt b}$ are higher-component operators, a non-vanishing r.h.s. in eqs. 
\eqref{wiS2} and \eqref{wiS3} signals that supersymmetry is broken in the corresponding vacuum.
The supersymmetric partner of these identities is the Ward identity involving the divergence of the stress tensor. This can be easily derived holographically by considering the invariance of the renormalized action under boundary diffeomorphisms, but we will not discuss it here.

%%%%%%%%%%%%%%%%%%%%%%%%%%%%%%%%%%%%%%%%%%%%
\subsection{Trace identities}
\label{HWI2}

In this section we derive the trace operator identities associated respectively with the energy-momentum tensor and the supercurrent.
Let us consider the latter first. From the $\e^-$ supersymmetry transformations \eqref{sources-e-}, and using \eqref{conversion}, 
for the variation of $S_\tx{ren}$ we get\footnote{As we pointed out already, there is a potential anomaly on the r.h.s. 
of this equation, as well as on the r.h.s. of \eqref{weyl-inv}. To compute these anomalies an explicit computation of the 
local counterterms $S_\tx{ct}$ is required. However, the anomalies only contribute ultralocal contact terms in 
the Ward identities, which are not relevant for the present discussion.} 
\begin{align}\label{epsminS}
\d_{\e^-}S_\tx{ren}= &\int d^4x\sqrt{-\wt\g}\(\frac{i}{2}\langle\ol{S}^{-\m}\wt\G_\m\rangle -\frac{9M}{\sqrt{2}}~\langle \ol{\cal O}^+_{\wt \z_b}\rangle\)e^{-\frac{8}{15}U}\e^-=0\ ,
\end{align}
which yields the following identity between the one-point functions of the gamma-trace of the supercurrent 
and of the operator ${\cal O}_{\wt \z^b}$ at non-zero sources and at the cut-off 
\be
\frac{i}{2}\langle\ol{S}^{-\m}\wt\G_\m\rangle =\frac{9M}{\sqrt{2}}~\langle \ol{\cal O}^+_{\wt\z_b}\rangle\,.
\ee
Again, from this identity one can compute relations between various correlation functions by further differentiating. Using the limits \eqref{QFT-1pt-fns}, we can remove the cut-off to obtain the relation 
\be
\label{sTraceWI}
\langle \sigma^\m_{\alpha\dot\beta} \bar S_\m^{\dot\beta} \rangle_\tx{QFT}=-9\sqrt{2} M~\langle {\cal O}_{\wt\z_b \,\alpha}\rangle_\tx{QFT}~.
\ee

Next, let us derive the Ward identity following from local shifts in the radial coordinate, which correspond to local Weyl transformations on the boundary. Using the transformation of the covariant sources given in eq. \eqref{weyl-transf}, we get  
\begin{align}\label{weyl-inv}
\d_{\s}S_\tx{ren}=& \int d^4x\sqrt{-\wt\g}\(\rule{0cm}{0.6cm}\frac12\d_\s\wt\g_{\m\n} \la T^{\m\n}\ra + 2\d_\s\f \, \la{\cal O}_\f\ra + 2\d_\s\wt b^\F \la{\cal O}_{\wt b}\ra\right.\NO\\
&\left.\hskip2.cm +\bigg[\frac i2\langle \ol{S}^{-\m} \rangle\d_\s\wt\J_\m^+-\sqrt{2}i \la \ol{\cal{O}}^+_{\z^\f}\ra\d_\s{\z}^{-}_\f - \sqrt{2}i\la \ol{\cal{O}}^+_{\wt\z^b}\ra\d_\s\wt\z^{-}_b + \tx{h.c.}\bigg] \)\NO\\
=&\int d^4x\sqrt{-\wt\g}\(\rule{0cm}{0.6cm} \la T^\m_\m\ra  + 9M \la{\cal O}_{\wt b}\ra\right.\\
&\left.\hskip2.cm +\bigg[\frac i4\langle \ol{S}^{-\m} \rangle \wt\J_\m^++\frac{i}{\sqrt{2}} \la \ol{\cal{O}}^+_{\z^\f}\ra {\z}^{-}_\f +\frac{i}{\sqrt{2}}\la \ol{\cal{O}}^+_{\wt\z^b}\ra\wt\z^{-}_b + \tx{h.c.}\bigg] \)e^{-\frac{8}{15}U}\s\ .\NO
\end{align}
 This leads to the following identity between bosonic one-point functions at the cut-off
\be
 \la T^\m_\m\ra  + 9M \la{\cal O}_{\wt b}\ra+\bigg[\frac i4\langle \ol{S}^{-\m} \rangle \wt\J_\m^++\frac{i}{\sqrt{2}} \la \ol{\cal{O}}^+_{\z^\f}\ra {\z}^{-}_\f +\frac{i}{\sqrt{2}}\la \ol{\cal{O}}^+_{\wt\z^b}\ra\wt\z^{-}_b + \tx{h.c.}\bigg]=0\,.
\ee
Removing the cut-off (and setting all sources to zero), we finally obtain
\be
\label{traceWI}
\la T^\m_\m\ra_\tx{QFT}= -9 M\la {\cal O}_{\wt b}\ra_\tx{QFT}\,.
\ee
This is the bosonic partner of the fermionic trace identity \eqref{sTraceWI} and the two are in perfect agreement, numerical coefficients included.

Notice that only the VEV of ${\cal O}_{\wt b}$ and not that of ${\cal O}_\f$ enters eq.~\eqref{traceWI}. 
From the general formula $T^\m_\m = - \frac 12 \sum_i \beta_i \, {\cal O}_i$ 
this suggests  that in the KS theory the operator ${\cal O}_\f$ remains marginal, at least in the 
supergravity regime, while $ {\cal O}_{\wt b}$ has non-trivial $\beta$-function. This is indeed the case, 
as shown in \cite{Klebanov:2000hb}, in perfect agreement  with the field theory answer in the large-$N$ limit. 
We will further comment on this point later.

%%%%%%%%%%%%%%%%%%%%%%%%%%%%%%%%%%%
%%%%%%%%%%%%%%%%%%%%%%%%%%%%%%%%%%%

\section{Bosonic one-point functions and the Goldstino}
\label{sec4}

Our goal here is to see how the supersymmetry Ward identities \eqref{wiS1}-\eqref{wiS3} are realized differently in the backgrounds \eqref{KTsol} and \eqref{nonSUSYbg}. 
Given the derivation of subsection \ref{HWI1}, it suffices to evaluate the bosonic one-point functions of $T_{\m\nu}, {\cal O}_\f$ and ${\cal O}_{\wt b}$.

The calculation of the bosonic VEVs in the background \eqref{nonSUSYbg} was already performed in \cite{Aharony:2005zr}. %\footnote{\Yiannis{Comment on different definition of one-point functions....}} 
The authors of that paper took the most general compactification of the normalizable deformations of the 10d KT solution. 
In particular, their solution contains transverse dependence and is obtained from 
an ansatz which is gauge-redundant because of radial diffeomorphisms. 
This makes the calculation of the VEVs technically involved. However, if we focus just on flat domain wall solutions and fix radial diffeomorphisms, we can obtain the one-point functions in a simpler manner.
With this simplification in mind, we provide below an independent derivation of the VEVs of $T_{\m\nu}, {\cal O}_\f$ and ${\cal O}_{\wt b}$, 
and find agreement with the results of \cite{Aharony:2005zr,DeWolfe:2008zy}.

In order to evaluate the bosonic one-point functions in \eqref{oneptf} explicitly, we compute separately the contributions coming respectively from the regularized action and the counterterms in \eqref{Sren}. The contribution coming from $S_\tx{reg}$ is the radial canonical momentum associated with the corresponding induced field, as follows from Hamilton-Jacobi theory.\footnote{As an elementary example consider the canonical momentum of a point particle described by the Lagrangian $L=\frac12 \dot x^2$, given by $p=\pa L/\pa\dot x=\dot x$. 
Invoking the equations of motion it follows that this canonical momentum can also be expressed as $p=\pa S_\tx{reg}/\pa x$,  where the on-shell action is identified with Hamilton's principal function.} Using the expressions for the radial canonical momenta corresponding to the fields $\wt\g_{\m\n}$, $\f$, and $\wt b^\F$ in the coordinate system 
(\ref{gf-conditions}) (see e.g. \cite{Papadimitriou:2004rz}) and using the identities \eqref{partial-derivatives}, the bosonic VEVs in \eqref{oneptf} become 
\begin{align}
\langle T^{\m\n}\rangle&=e^{\frac{4}{15}U}e^{X}\left(- 2 \left(K\g^{\m\n}-K^{\m\n}\right)+\frac{2}{\sqrt{-\g}}\frac{\d S_{\rm ct}}{\d\g_{\m\n}}\right)\ , \label{Tvev}\\
\langle {\cal O}_\f\rangle &=- e^{X}\({\cal G}_{\f\f}\dot\f+b^\F\cg_{b^\F b^\F}\dot b^\F +\left(1+\frac k2e^{-\frac45 U}\right)\left(\frac54\cg_{UU}\dot U-\frac12K\right)\)\NO\\
&\hskip0.5cm+\frac{1}{2\sqrt{-\g}}e^{X}\left(\frac{\d S_{\rm ct}}{\d\f}+b^\F\frac{\d S_{\rm ct}}{\d b^\F}+\left(1+\frac k2e^{-\frac45 U}\right)\left(\frac54\frac{\d S_{\rm ct}}{\d U}+\frac13\g_{\m\n}\frac{\d S_\tx{ct}}{\d\g_{\m\n}}\right)\right),\label{Ovev1} \\
\langle {\cal O}_{\wt b}\rangle &=e^\f e^{X}\left(- {\cal G}_{b^\F b^\F}\dot b^\F-\frac{q}{2}e^{-\frac45U}\(\frac54\cg_{UU}\dot U-\frac12K\)\right.\NO\\
&\hskip0.5cm\left.+\frac{1}{2\sqrt{-\g}}\left(\frac{\d S_{\rm ct}}{\d b^\F}+\frac{q}{2}e^{-\frac45U}\(\frac54\frac{\d S_{\rm ct}}{\d U}+\frac13\g_{\m\n}\frac{\d S_\tx{ct}}{\d\g_{\m\n}}\)\right)\right),\label{Ovev2}
\end{align}
where the dot represents derivatives with respect to the radial coordinate $r$, which is defined in eq. (\ref{gf-conditions}), while $K_{\m\n}$ 
is the extrinsic curvature of the radial slices which, for the metric \eqref{gf-conditions}, takes the form
\be\label{excurvature}
K_{\m\n}={1\over 2}\dot\g_{\m\n}=-{1\over 2}ze^{-X}\del_z\left(\frac{e^{2Y}}{z^2}\right)\eta_{\m\n}\ .
\ee
The contribution to the bosonic VEVs from $S_\tx{ct}$ requires to know the explicit form of the (bosonic part of the) boundary counterterms, at least for the case of Poincar\'e domain wall solutions. Both the bosonic and fermionic counterterms can be derived systematically for general cascading solutions. 
For backgrounds enjoying 4D Poincar\'e invariance it turns out that the bosonic counterterms in a supersymmetric scheme \cite{Bianchi:2001de} are simply given by the superpotential \eqref{sup}, namely   
\be
S_\tx{ct}=-\int d^4x \sqrt{-\g}\;2\cw\ .
\ee
Putting the two contributions together, the VEVs (\ref{Tvev})-(\ref{Ovev2}) at the radial cut-off take the form 
\begin{align}
\langle T^\m_\n\rangle &= -2 \left[3z\del_z\log\left(\frac{e^{Y}}{z}\right)+e^{X}{\cal W}\right]\d^\m_\n\ ,\nn\\
\langle {\cal O}_\f\rangle&=  \frac12z\del_z\f+e^{-\f}b^\F\left(e^{-{4\over 5}(U+V)}z\del_zb^\F-e^\f e^{X}\del_{b^\F}{\cal W}\right)\NO\\
&\hskip0.5cm+\left(1+\frac k2e^{-\frac45 U}\right)\[\frac54\left(\frac{8}{15}z\pa_z U-e^{X}\pa_U\cw\right)-2z\pa_z\log\left(\frac{e^Y}{z}\right)-\frac23 e^X\cw\]\ ,\NO\\
\langle {\cal O}_{\wt b}\rangle&= e^{-{4\over 5}(U+V)}z\del_zb^\F-e^\f e^{X}\del_{b^\F}{\cal W}\NO\\
&\hskip0.5cm+\frac{q}{2}e^{-\frac45 U+\f}\[\frac54\left(\frac{8}{15}z\pa_z U-e^{X}\pa_U\cw\right)-2z\pa_z\log\left(\frac{e^Y}{z}\right)-\frac23 e^X\cw\]~.
\end{align}
Evaluating the limits in \eqref{QFT-1pt-fns} using the asymptotic behavior of the induced fields we finally get 
\begin{align}
\label{vE}
\langle T^\m_\m \rangle_\tx{QFT}&= -12 \, {\cal S}~, \\
\langle {\cal O}_\f\rangle_\tx{QFT}&= \frac{(3{\cal S}+4\varphi)}{2}\ ,\\
\label{vOb}
\langle {\cal O}_{\wt b}\rangle_\tx{QFT}&=  \frac{4}{3 \, M}{\cal S}~,
\end{align}
in agreement with the corresponding expressions in \cite{Aharony:2005zr,DeWolfe:2008zy} (note that the sign of ${\cal S}$ is univocally fixed from \eqref{vE}, by unitarity). 

Let us elaborate on the above result, utilizing the Ward identities \eqref{wiS1}-\eqref{wiS3}, 
which we derived holographically. On the supersymmetric solution \eqref{KTsol}, for which ${\cal S}=\varphi=0$,
all the above VEVs vanish, i.e. the vacuum energy is zero and there are no non-trivial VEVs 
for higher component operators. The supersymmetry Ward identities are trivially satisfied, and there is no massless 
pole in the supercurrent two-point function \eqref{sbars}. This is all consistent with supersymmetry being preserved. 

More interestingly, let us now look at non-supersymmetric branches, and start with the branch ${\cal S}=0 ,\; \varphi \not=0$. Here supersymmetry is broken in the dual field theory, since a higher-component operator, ${\cal O}_\f$,  has a non-vanishing VEV. 
Since $\la T^\m_\m \ra=0$, however, the vacuum energy vanishes and the goldstino mode is absent in \eqref{sbars}. 
This is an indication of explicit supersymmetry breaking, meaning that this branch does not describe vacua of the KS model. 
This agrees with the fact that the $\beta$-function  of the sum of the inverse gauge coupling squared, the coupling dual to 
${\cal O}_\f$, actually vanishes \cite{Klebanov:2000hb} and hence ${\cal O}_\f$ remains exactly marginal. 
As such, it  cannot trigger spontaneous supersymmetry breaking (the dynamics along this branch is basically the 
same as in the case of the dilaton background of \cite{Gubser:1999pk,Kehagias:1999tr}, though in a non-conformal theory).

Finally, let us consider the branch $\varphi = 0,\; {\cal S} \not =0$. This was suggested in \cite{DeWolfe:2008zy} to correspond 
to the (asymptotic description of the) metastable state obtained by placing $p \sim {\cal S}$ antiD3-branes at the tip of the deformed conifold. 
Along this branch we see that the vacuum energy \eqref{vE} is non-vanishing, this being triggered by the VEV of the operator ${\cal O}_{\wt b}$, eq.~\eqref{vOb}. 
Indeed, these two quantities exactly satisfy the relation $T^\m_\m = -\frac 12 \beta_{\wt b^\F} {\cal O}_{\wt b}$ (the difference with respect to the normalization of \cite{Klebanov:2000hb} is just due to a different normalization of the operator ${\cal O}_{\wt b}$).  
From the supercurrent Ward identities \eqref{wiS1} and \eqref{wiS3}, which hold non-trivially in this vacuum, we see that a goldstino mode 
is present in the supercurrent two-point function \eqref{sbars}. From the operator identity \eqref{sTraceWI} it follows that the goldstino 
eigenstate is\footnote{It is worth noticing that, from a field theory viewpoint, 
there are no obvious symmetries protecting the dimension of ${\cal O}_\f$. 
Hence, one would expect its dimension to get corrections, at least beyond the supergravity regime. 
Evidence for this was given in \cite{Frolov:2001xr}, where $\alpha'^3$-corrections were computed 
suggesting that the otherwise marginal operator gets contributions to its anomalous dimension at order 
$\sim (\frac{M}{N})^4 (g_s N)^{-1/2}$ (recall that the supergravity limit is $g_s N \rightarrow \infty$). 
So, given that in this branch $\langle {\cal O}_\f\rangle_\tx{QFT} \not =0$, the goldstino eigenstate could 
get a (very much suppressed) contribution from ${\cal O}_{\z_\f}$, too, in the KPV vacuum. We thank Igor Klebanov for a discussion on this point.}
\be
G \sim \langle \co_{\wt b} \rangle \, \sigma^\m \bar S_\m \sim  \langle \co_{\wt b} \rangle \, \co_{\wt\z_{b}}\,.  
\ee
All these properties are consistent with a vacuum where supersymmetry is spontaneously broken and suggest that 
(if it exists, cf. the discussion in the Introduction) the KPV vacuum is in fact a vacuum of the KS theory.

%%%%%%%%%%%%%%%%%%%%%%%%%%%%%%%%%%%%%%%%%%%%%%
%%%%%%%%%%%%%%%%%%%%%%%%%%%%%%%%%%%%%%%%%%%%%%
\section{Conclusions}

The main focus of this paper was to derive holographically the supersymmetry Ward identities of the conifold cascading gauge theory, 
and to evaluate them explicitly in supersymmetric and supersymmetry-breaking dual backgrounds. 
Within the consistent truncation we have considered, a two-parameter family of supersymmetry-breaking solutions exists  with the correct asymptotics. 
We have shown that only a one-dimensional branch respects the supersymmetry Ward identities and displays the expected goldstino mode. 
This branch was conjectured in \cite{DeWolfe:2008zy} to describe, asymptotically, the state constructed by placing  antiD-branes at the tip of the deformed 
conifold, which is a metastable state in the probe approximation \cite{Kachru:2002gs}. In this sense, our results provide evidence that the KS 
cascading theory can admit vacua where supersymmetry is broken at strong coupling, and also that antiD-brane states, 
if they exist beyond the probe approximation, are valuable candidates for such vacua.\footnote{It would be interesting to repeat our computation for the solution of \cite{{Bena:2011wh}}, which includes also the conifold deformation parameter. The computation is more involved, since the truncation one should consider includes more fields. However, as already argued, we do not expect any qualitative changes in the end result.}

The derivation of the supersymmetry Ward identities we performed is quite general and does not rely very much on the specific structure of the conifold theory, 
nor on the explicit form of the solutions. This suggests that supersymmetry breaking vacua might be generic in quiver gauge theories  
with running couplings driven by fractional branes, the KS model being just a prototype example 
(superconformal theories cannot break supersymmetry spontaneously, hence fractional branes are a necessary ingredient in the construction). 
Considering this larger class of theories, in terms of more general 5d sigma-models than the one presented in appendix \ref{sigma-model}, could be instructive.\footnote{In the probe approximation, where the goldstino is a massless excitation on the antiD3-brane world-volume, this was shown to be the case in generalizations of the KPV construction on conifold-like geometries with orientifolds \cite{Kallosh:2015nia}, see also \cite{Bergshoeff:2015jxa,Ferrara:2014kva}.}

Our results are consistent with previous findings \cite{Aharony:2005zr,Buchel:2005cv,Aharony:2006ce}, where it was suggested that 
cascading theories, although being rather unconventional from the field theory point of view, are in fact renormalizable holographically (see also \cite{Berg:2005pd,Berg:2006xy,Haack:2010zz}). 
There are however several remaining open questions. The derivation of the counterterms we pursued is all one needs to renormalize bosonic 
one-point functions, but this is not the full story. In fact, the approach we used, where correlators are defined in terms of induced fields at a finite cut-off rather than in terms of sources, seems robust and general enough to let one compute the full counterterm action, including all bosonic and fermionic counter-terms. This could make the analysis 
initiated in \cite{Aharony:2005zr,Aharony:2006ce} more rigorous and possibly far reaching.\footnote{We thank Amos Yarom for a discussion on this point.} 
Work is in progress in this direction \cite{usfut}. 

Working in terms of induced fields looks also as an efficient approach to try and answer the question on how to derive,
from first principles, counterterms respecting supersymmetry in generic setups.  In fact, this could also provide a technically 
and conceptually promising way to attack the problem of holographically renormalize supersymmetric theories on curved manifolds. 
We hope to return on this issue in the near future.

%%%%%%%%%%%%%%%%%%%%%%%%%%%%%%%%%%%%%%%%%%%%%%
\section*{Acknowledgements}

We are grateful to Riccardo Argurio, Giulio Bonelli, Lorenzo Di Pietro, Anton Faedo, Gabriele Ferretti, Igor Klebanov, Stefano Massai, Flavio Porri, 
Diego Redigolo, Marco Serone, Phillip Szepietowski, Tomas Van Riet, and Amos Yarom for discussions and/or email exchanges.
D.M. thanks also Andrea Amoretti, Victor Giraldo Rivera, Rajesh Gupta and Leopoldo Pando Zayas for interesting discussions.

%%%%%%%%%%%%%%%%%%%%%%%%%%%%%%%%%%%%%%%%%%%%%%%%%%%%%

\appendix

%%%%%%%%%%%
\section{The 5d supergravity action}
\label{sigma-model}
In this appendix we collect all relevant details of the five-dimensional supergravity theory we work with. The theory we consider was obtained in \cite{Cassani:2010na,Bena:2010pr,Liu:2011dw,Halmagyi:2011yd} as an ${\cal N}=2$ consistent truncation of the $SU(2) \times SU(2)$ invariant sector of Type IIB supergravity on $\T$. In fact, as explained in the main text, we focus on a truncation preserving an additional $U(1)$ symmetry \cite{Buchel:2014hja}.

The bosonic action, restricted to the fields relevant for our analysis, namely the metric $g_{\m\n}$ and the four scalars $U, V, b^\F$ and $\f$, can be written as a $\sigma$-model and reads
\be
\label{sigma}
S_b=\frac{1}{2\k^2}\int d^{5}x\sqrt{-g}\,\bigg(R-\cg_{IJ}(\vf)\pa_A\vf^I\pa^A\vf^J-{\cal V}(\vf)\bigg)~.
%%%%%%%%%%%%%%%%%%%%%%%%%%%%%%%%%%%%%%%%%%%%%%%%%%%%%%%%%%%%%
\ee
The fermionic action containing the gravitino $\J_M$ and the four spinor fields $\z_U$, $\z_V$, $\z_{b^\F}$ and $\z_\f$ can also be expressed in terms of sigma model language and, up to quadratic terms in the fermions takes the form
\bea
&S_f=- \frac{1}{2\k^2} \int d^{5}x\sqrt{-g}\,\bigg[\frac12\left(\lbar{\J}_A\G^{ABC}{\cal D}_B\J_C+i\cg_{IJ}\lbar\z^I\G^A\left(\slashed\pa\vf^J-\cg^{JK}\pa_K\cw\right)\J_A+\tx{h.c.}\right)\nn\\
&\hskip3.0cm+\frac12\left(\cg_{IJ}\lbar\z^I\left(\d^J_K\slashed\nabla+\G^{J}_{KL}[\cg]\slashed\pa\vf^L\right)\z^K+h.c.\right)+\cm_{IJ}(\vf)\lbar\z^I\z^J\bigg]~.\label{Sf}
\eea
Here, $\k^2=8\p G_5$ and the indices $A,B,...$ are 5d space-time indices, while $I,J,...$ are indices on the scalar manifold. In particular, 
\be\label{scalarmetric}
\vf^I=\left(\begin{array}{c}
U\\ V \\ b^\F \\ \f 
\end{array}\right),\quad \z^I=\left(\begin{array}{c}
\z_U\\ \z_V\\ \z_b\\ \z_\f 
\end{array}\right),\quad \cg_{IJ}(\vf)=\left(
\begin{array}{cccc}
 \frac{8}{15} & 0 & 0 & 0 \\
 0 & \frac{4}{5} & 0 & 0 \\
 0 & 0 & e^{-\frac{4}{5} (U+V)-\f} & 0 \\
 0 & 0 & 0 & \frac{1}{2} \\
\end{array}
\right).
\ee
The only non-zero components of the Christoffel symbol $\G^{K}_{IJ}[\cg]$ of the metric \eqref{scalarmetric} on the scalar manifold are
\bal
&\G^{U}_{b^\F b^\F}[\cg]=\frac34e^{-\frac45(U+V)-\f},\quad \G^{V}_{b^\F b^\F}[\cg]=\frac12e^{-\frac45(U+V)-\f},\quad \G^{\f}_{b^\F b^\F}[\cg]=e^{-\frac45(U+V)-\f},\NO\\ 
&\G^{b^\F}_{b^\F U}[\cg]=-\frac25,\quad \G^{b^\F}_{b^\F V}[\cg]=-\frac25,\quad \G^{b^\F}_{b^\F \f}[\cg]=-\frac12~. 
\eal
The covariant derivative $\nabla_A$ and the supercovariant derivative ${\cal D}_A$ are defined as follows
\begin{subequations}
\begin{align}
\nabla_A & = \del_A+{1\over 4}\(\omega_A\)^{ab}\g_{ab}\ ,\\
{\cal D}_A &= \nabla_A +{1\over 6} \G_A {\cal W}\ ,
\end{align}
\end{subequations}
where $a,b,...$ are indices on the tangent space and $\(\omega_A\)^{ab}$ is the spin connection of the 5d metric.
The scalar potential takes the following form
\be
{\cal V}(\vf)=2 e^{-{8\over 3}U}(b^\F q-k)^2+ e^{-{4\over 15}(7U-3V)+\f}q^2-24 e^{-{2\over 15}(8U+3V)}+4 e^{-{4\over 15}(4U+9V)}~,
\ee
where we used the following relations to connect to the notations adopted in the main text
\be\label{conversion}
q=\frac{9}{2}M~,\quad k=-\frac{27\pi N}{2}~,
\ee
with $N$ and $M$ being the number of regular and fractional D3 branes respectively.
Both the scalar potential and the mass matrix $\cm_{IJ}$ can be expressed in terms of the Papadopoulos-Tseytlin superpotential \cite{Papadopoulos:2000gj}
\be\label{sup}
{\cal W}(\vf)=(k-qb^\F)e^{-{4\over 3}U}+3e^{-{4\over 15}(2U-3V)}+2e^{-{2\over 15}(4U+9V)}~,
\ee
through the relations
\begin{subequations}
\bal\label{sup-eq}
\cv(\vf)=&\, \cg^{IJ}\pa_I{\cal W}(\vf)\pa_J{\cal W}(\vf)-\frac{4}{3}{\cal W}(\vf)^2\ ,\\
\cm_{IJ}(\vf)=&\, \pa_I\pa_J\cw-\G^{K}_{IJ}[\cg]\pa_K\cw-\frac12\cg_{IJ}\cw\ .
\eal
\end{subequations}
The supersymmetry transformations to linear oder in $\e$ are
\begin{subequations}
\bal
\d_\e\z^I=&\, -\frac i2\left(\slashed\pa\vf^I-\cg^{IJ}\pa_J\cw\right)\e\ ,\label{ezeta}\\
\d_\e\J_A=&\,\left(\nabla_A+\frac16\cw\G_A\right)\e\ ,\label{gravitinovar}\\
\d_\e\vf^I=&\,{i\over 2} \bar\e\z^I+\tx{h.c.}\ ,\\
\d_\e e^a_A=&\,\frac12\bar \e\G^a\J_A+\tx{h.c.}\ .
\eal
\end{subequations}
It follows that the BPS equations for Poincar\'e domain wall solutions of the form \eqref{ansa} are
\be\label{BPSeqs}
e^{-X(z)}z\pa_z\vf^I-\cg^{IJ}\pa_J\cw=0\ ,\quad e^{-X(z)}z\pa_z\log\left(\frac{e^Y}{z}\right)+\frac13\cw=0\ .
\ee

%%%%%%%%%%%%%%%%%%%%%%%%%%%%%
\section{Equations of motion and leading asymptotics}
\label{eoms}

In this appendix we give the bosonic and fermionic equations of motion following from the action \eqref{sigma}+\eqref{Sf}, as well as the leading form of the their asymptotic solutions, subject to KT boundary conditions.  

\subsection{Bosonic sector}

In the bosonic sector the equations of motion are
\begin{subequations}
\bal
&\label{EQdilaton}
\frac{1}{\sqrt{-g_5}}\del_{A}\(\sqrt{-g_5}~g^{AB}~\del_{B}\f\)=e^{-{4\over 15}(7U-3V)+\f}q^2- e^{-{4\over 5}(U+V)-\f}\(db^\F\)^2\ ,\\\NO\\
&\label{EQbphi}
\frac{1}{\sqrt{-g_5}}\del_{A}\(\sqrt{-g_5}~e^{-{4\over 5}(U+V)-\f}~g^{AB}~\del_{B}b^\F\)=2q~e^{-{8\over 3}U}\(b^\F q-k\)\ ,\\\NO\\
&\label{EQU}
{16\over 15}\frac{1}{\sqrt{-g_5}}\del_{A}\(\sqrt{-g_5}~g^{AB}~\del_{B}U\)+{4\over 5}e^{-{4\over 5}(U+V)-\f}
\(db^\F\)^2 +{16\over 3}e^{-{8\over 3}U}(b^\F q-k)^2\nn\\
& +{28\over 15}e^{-{4\over 15}(7U-3V)+\f}q^2+{64\over 15}e^{-{4\over 15}(4U+9V)}-{128\over 5}e^{-{2\over 15}(8U+3V)}=0\ , \\\NO\\
&\label{EQV}
{8\over 5}\frac{1}{\sqrt{-g_5}}\del_{A}\(\sqrt{-g_5}~g^{AB}~\del_{B}V\)+{4\over 5}\bigg(e^{-{4\over 5}(U+V)-\f}
\(db^\F\)^2-q^2e^{-{4\over 15}(7U-3V)+\f}\bigg)\nn\\
&+{48\over 5}\bigg(e^{-{4\over 15}(4U+9V)}-e^{-{2\over 15}(8U+3V)}\bigg)=0\ ,\\\NO\\
&R_{AB}={8\over 15} \del_A U\del_B U +{4\over 5}  \del_A V\del_B V  +e^{-{4\over 5}(U+V)-\f}
\del_A b^\F\del_B b^\F+{1\over 2} \del_A\f\del_B\f\nn\\
&+{1\over 3}g_{AB}\bigg(2 e^{-{8\over 3}U}(b^\F q-k)^2+ e^{-{4\over 15}(7U-3V)+\f}q^2
-24 e^{-{2\over 15}(8U+3V)}
+4 e^{-{4\over 15}(4U+9V)}\bigg)\ .
\eal
\end{subequations}

\subsubsection*{Asymptotic solutions}

In order to obtain the asymptotic solutions of the equations of motion it is necessary to pick a specific gauge. In the gauge \eqref{gf-conditions}, the leading asymptotics of the bosonic fields for any solution that asymptotes to the KT solution take the form  
\bal\label{b-asymptotics}
&\g_{\m\n}(z,x)\sim \frac{h^{1/3}(z)}{z^2}\(\left(1  +\frac{1}{24}g_sq^2h^{-1}(z)(1-4\log z)c(x)+\frac16qh^{-1}(z)~b(x)\right)\h_{\m\n}+h_{\m\n}(x)\),\NO\\
&\f(z,x) \sim \log g_s+c(x)\ ,\NO\\
&b^\F(z,x) \sim b(x)-(1+c(x)) ~g_s q\log z\ ,\NO\\
&U(z,x) \sim {5\over 4}\log \(h(z)+{1\over 8}g_sq^2(1-4\log z)c(x)+{1\over 2}q~b(x)\)\ ,\NO\\
&V(z,x) = {\cal O}(z^4)\ ,
\eal
where the warp factor is given by
\be\label{WarpF}
h(z)={1\over 8}\(-4k+g_sq^2-4g_s q^2\log z\) +{\cal O}(z^4),
\ee
and $h_{\m\n}(x),c(x)$ and $b(x)$ are infinitesimal sources.

\subsection{Fermionic sector}

The fermionic equations of motion take the form
\begin{subequations}
\label{f-eoms}
\bal
&\label{dilatino}
\slashed{\nabla}\z_\f+\frac i2\G^M\slashed\pa\f\J_M-m_\f\z_\f+{\cal F}_-{\cal G}_{b^\F b^\F}^{1/2}\z_b=0\ ,\\\NO\\
&\slashed{\nabla}\({\cal G}_{b^\F b^\F}^{1/2}\z_b\)+\frac i2\G^M\cf_+\J_M+m_b{\cal G}_{b^\F b^\F}^{1/2}\z_b-{1\over 2}{\cal F}_+\z_\f+\del_U{\cal F}_-\z_U+\del_V{\cal F}_-\z_V=0\ ,\\\NO\\
&\slashed{\nabla}\z_U+\frac i2\G^M\cb_+^U\J_M+m_U\z_U+{12\over 5}\(e^{-{2\over 15}(4U+9V)}-e^{-{4\over 15}(2U-3V)}\)\z_V-{15\over 8}\del_U{\cal F}_+{\cal G}_{b^\F b^\F}^{1/2}\z_b=0\ ,\\ %\NO\\
&\slashed{\nabla}\z_V+\frac i2\G^M\cb_+^V\J_M+m_V\z_V+{8\over 5}\(e^{-{2\over 15}(4U+9V)}-e^{-{4\over 15}(2U-3V)}\)\z_U-{5\over 4}\del_V{\cal F}_+{\cal G}_{b^\F b^\F}^{1/2}\z_b=0\ ,\\ %\NO\\
&\G^{ABC}{\cal D}_B\J_C -{i\over 2}\({1\over 2}\slashed{\del}\f\G^A\z_\f+{\cal F}_-\G^A{\cal G}_{b^\F b^\F}^{1/2}\z_b+{8\over 15}{\cal B}^U_-\G^A\z_U+{4\over 5}{\cal B}^V_-\G^A\z_V\)=0\ ,
\eal
\end{subequations}
where we have defined the following quantities
\begin{subequations}
\begin{align}
{\cal F}_\pm&={\cal G}_{b^\F b^\F}^{1/2}\(\slashed{\del}b^\F\pm e^{-{4\over 15}(2U-3V)+\f}q\)\ ,\\
{\cal B}_\pm^U&=\slashed{\del}U\pm{1\over 2}\(5(k-b^\F q)e^{-{4\over 3}U}+6e^{-{4\over 15}(2U-3V)}+4e^{-{2\over 15}(4U+9V)}\)\ ,\\
{\cal B}_\pm^V&=\slashed{\del}V\pm 3\(e^{-{2\over 15}(4U+9V)}-e^{-{4\over 15}(2U-3V)}\)\\
m_\f(\vf)&={1\over 2}{\cal W}\ ,\\
m_b(\vf)&={1\over 2}{\cal W}-3e^{-{4\over 15}(2U-3V)}\ ,\\
m_U(\vf)&={1\over 30}\({\cal W}+84\(k-b^\F q\)e^{-{4\over 3}U}\)\ ,\\
m_V(\vf)&={3\over 10}{\cal W}-{4\over 5}(k-b^\F q)e^{-{4\over 3}U}+2e^{-{2\over 15}(4U+9V)}\ . 
\end{align}
\end{subequations}
The fermion masses $m_\f(\vf),\, m_b(\vf),\, m_U(\vf),\, m_V(\vf)$ 
reproduce the masses shown in Table \ref{massSpectrum} in the AdS limit $(q\to 0)$ with unit AdS radius $(k= - 2)$.

\subsubsection*{Asymptotic solutions}

In the gauge \eqref{gf-conditions} the leading asymptotics of the fermions, for any bosonic solution that asymptotes to the KT solution, take the form
\bal\label{f-asymptotics}
&\J_\m^+(z,x)\sim z^{-1/2}h(z)^{1/12}\J_{(0)\m}^+(x)\NO\\
&\hskip2.cm+iz^{-1/2}h(z)^{1/6}\g_\m\(-\frac{4}{5g_s q^2}h(z)^{11/12}\psi_1^-(x)+ \frac{h(z)^{-7/4}}{12 g_s q}\(g_s q^2+12 h(z)\)\psi_2^-(x)\)\ ,\NO\\
&\z_\f^-(z,x) \sim z^{1/2}h(z)^{-1/12}\j_1^{-}(x)\ ,\NO\\
&\z_b^-(z,x) \sim \frac{z^{1/2}h(z)^{-1/12}}{20 q}\(24 h(z)-5g_s q^2\)\j_1^-(x)+z^{1/2}h(z)^{-3/4}\j_2^-(x)\ ,\NO\\
&\z_U^-(z,x) \sim \frac{3}{4}z^{1/2}h(z)^{-1/12}\j_1^-(x)+\frac{5}{8}q~ z^{1/2}h(z)^{-7/4}\j_2^-(x)\ ,\NO\\
&\z_V^+(z,x)= {\cal O}(z^{3/2})\ ,
\eal
where $h(z)$ is given in \eqref{WarpF} and $\J_{(0)\m}^+(x)$, $\j_1^{-}(x)$, $\j_2^{-}(x)$ are spinor sources of the indicated chirality. Notice that the limit $q\to 0$, corresponding to KW asymptotics, is a singular limit in these asymptotic solutions. In particular, the parameter $q$ corresponds to a singular perturbation of the fermionic equations of motion \eqref{f-eoms}.

%%%%%%%%%%%%%%%%%%%%%%%%%%%%%%%%%%%%%%%%%%%%%%%%%%%%%

\section{Covariant sources for gauge-invariant operators}
\label{hol-map}

As was mentioned in section \ref{HWI}, the covariant sources of certain operators in the KS theory are composite in terms of bulk fields. In particular, the covariant source of the difference of the inverse gauge couplings square corresponds to the composite field $\wt b^\F=e^{-\f}b^\F$. Inserting the asymptotic expansions \eqref{b-asymptotics} we find that $\wt b^\F$ asymptotes to   
\be
\wt b^\F\sim  g_s^{-1}b(x)-q\log z\ ,
\ee
and it is therefore sourced only by the $b(x)$ mode. Similarly, the composite field 
\be
\wt U= \frac18\left(4q b^\F-4k+q^2 e^\f\right)e^{-\frac45 U}\ ,
\ee
has the property that the modes $b(x)$ and $c(x)$ drop out of its asymptotic expansion so that $\wt U=1$, up to normalizable modes. Moreover, the BPS equations \eqref{BPSeqs} imply that $\wt U$ is a constant, up to a mode that has the right scaling to be identified with the source of a dimension 8 operator, which therefore corresponds to a supersymmetric irrelevant deformation \cite{Gubser:1999pk,Intriligator:1999ai}. These two properties allow us to identify $\wt U$ with the covariant source of the dimension 8 operator, which can therefore be consistently switched off by setting $\wt U=1$. Finally, although this is not necessary, it is natural to define the stress tensor as the operator that couples only to the fluctuation $h_{\m\n}$ in \eqref{b-asymptotics}, which can be achieved by defining the covariant source of the stress tensor as
\be
\wt\g_{\m\n}=e^{-\frac{4}{15}U}\g_{\m\n}\ .
\ee 

The covariant sources for the fermionic partners of these operators follow by supersymmetry and are given respectively by\footnote{In fact, the fully covariant with respect to $\wt\g_{\m\n}$ fermionic sources contain an additional factor of $e^{U/15}=e^{X/8}$, which comes from the covariantization of the spinor $\e^+$ with respect to $\wt\g_{\m\n}$. This extra factor would remove the factors of $e^{-X/8}$ from the definition of the fermionic one-point functions in \eqref{QFT-1pt-fns}, as well as an overall factor of $h^{-1/12}$ from the expansions \eqref{f-source-expansions}. However, since we are working to linear order in the sources this factor does not play a crucial role and we have chosen not to include it in the definition of the fermion sources. }
\bal
&\wt\z_b^-=e^{-\f}\left(\z_b^--b^\F\z_\f^-\right)\ ,\NO\\
&\wt\z_U^-=-\frac 45 \wt U\z^-_U+\frac18e^{-\frac45 U}\left(4q\z_b^-+q^2e^\f\z_\f^-\right)\ ,\NO\\
&\wt\J_\m^+=e^{-\frac{2}{15}U}\left(\J_\m^+-\frac{2i}{15}\G_\m\z_U^-\right)\ .
\eal
The leading asymptotic behavior of these fields, following from \eqref{b-asymptotics} and \eqref{f-asymptotics}, is  
\bal\label{f-source-expansions}
&\wt\z_b^-\sim -\frac{1}{5g_sq}z^{1/2}h(z)^{-1/12}\(4 h(z)+5k\)\j_1^-(x)+g_s^{-1}z^{1/2}h(z)^{-3/4}\j_2^-(x)\ ,\NO\\
&\wt\z_U^-\sim 0\ ,\\
&\wt\J_\m^+\sim z^{-1/2}\left(h(z)^{-1/12}\left(\J_{(0)\m}^+-\frac{i}{10}\g_\m\j_1^-\right)-\frac{4i}{5g_sq^2}h(z)^{11/12}\g_\m\j_1^-+\frac{i}{g_sq}h(z)^{-3/4}\g_\m\j_2^-\right)\ ,\NO
\eal
where $\wt\z_U^-$ is only sourced by a mode corresponding to an irrelevant operator of dimension $15/2$, which can therefore be put to zero consistently.

The fact that the covariant sources $\wt\g_{\m\n}$, $\wt b^\F$ and $\wt U$, as well as their supersymmetric partners, are composite in terms of supergravity fields implies that some care is required when evaluating the partial derivatives in the definition of the one-point functions \eqref{oneptf}, where composite fields are held constant. In particular, expressing the supergravity fields in terms of the composite fields, 
\bal
&b^\F=e^\f\wt b^\F\ ,\NO\\
&e^{\frac45U}=\frac18\wt U^{-1}e^\f\left(4q\wt b^\F-4k e^{-\f}+q^2\right)\ ,\NO\\
&\g_{\m\n}=\frac12\wt U^{-1/3}e^{\f/3}\left(4q\wt b^\F-4k e^{-\f}+q^2\right)^{1/3}\wt\g_{\m\n}\ ,\NO\\
&\z_b^-=e^\f\left(\wt\z_b^-+\wt b^\F\z_\f^-\right)\ ,\NO\\
&\z_U^-=-\frac54\wt U^{-1}\left(\wt \z_U^--\wt U \left(4q\wt b^\F-4k e^{-\f}+q^2\right)^{-1}\left(4q\wt\z_b^-+(4q\wt b^\F+q^2 )\z_\f^- \right)\right)\ ,\NO\\
&\J_\m^+=\frac{1}{\sqrt{2}}\wt U^{-1/6}e^{\f/6}\left(4q\wt b^\F-4k e^{-\f}+q^2\right)^{1/6}\left(\wt\J_\m^++\frac{2i}{15}\wt\G_\m\z_U^-\right)\ ,
\eal
one obtains the following expressions for the partial derivatives of a generic function $F$ with respect to the covariant bosonic sources 
\bal\label{partial-derivatives}
\left.\frac{\pa F}{\pa\wt\g_{\m\n}}\right|_{\f,\wt b^\F,\wt U,\wt\J^+,\z_\f^-,\wt\z_b^-,\wt\z_U^-}&=e^{\frac{4}{15}U}\frac{\pa F}{\pa\g_{\m\n}}\ ,\NO\\
% % % % % % % % % % % % % % % % % % % % % % % % % % % % % % % % % %
\left.\frac{\pa F}{\pa\f}\right|_{\wt\g,\wt b^\F,\wt U,\wt\J^+,\z_\f^-,\wt\z_b^-,\wt\z_U^-}&=\frac{\pa F}{\pa\f}+b^\F\frac{\pa F}{\pa b^\F}+\left(1+\frac k2e^{-\frac45 U}\right)\left(\frac54\frac{\pa F}{\pa U}+\frac13\g_{\m\n}\frac{\pa F}{\pa\g_{\m\n}}\right)+\tx{fermions}\ ,\NO\\
% % % % % % % % % % % % % % % % % % % % % % % % % % % % % % % % % % 
\left.\frac{\pa F}{\pa\wt b^\F}\right|_{\wt\g,\f,\wt U,\wt\J^+,\z_\f^-,\wt\z_b^-,\wt\z_U^-}&=e^\f\frac{\pa F}{\pa b^\F}+\frac{q}{2}e^{-\frac45 U+\f}\left(\frac54\frac{\pa F}{\pa U}+\frac13\g_{\m\n}\frac{\pa F}{\pa\g_{\m\n}}\right)+\tx{fermions}\ .
\eal
These expressions are required in order to correctly evaluate the one-point functions \eqref{oneptf}.

%%%%%%%%%%%%%%%%%%%%%%%%%%%%%
\section{Local symmetries and transformation of the sources}
\label{local-symmetries}

The bulk equations of motion dictate that certain components of the metric and of the gravitino are non-dynamical. In particular, the radial-radial and radial-transverse components of the metric (or, more precisely, the shift and lapse functions of the metric with respect to the radial coordinate), as well a the radial component of the gravitino, are non-dynamical and can be gauge-fixed to a convenient choice. We choose the gauge   
\be\label{gf-conditions}
ds_5^2=dr^2+\g_{\m\n}(r,x)dx^\m dx^\n,\quad \J_r=0\ ,
\ee
where the canonical radial coordinate $r$ is related to the coordinate $z$ in \eqref{ansa} through $dr=-e^{X(z)}dz/z$. Moreover, for the domain wall ansatz in \eqref{ansa} we have
\be
\g_{\m\n}=\frac{e^{2Y}}{z^2} \eta_{\m\nu}\ .
\ee
The gauge-fixing conditions \eqref{gf-conditions} are preserved by a subset of bulk diffeomorphisms and supersymmetry transformations. The transformation of the covariant sources under these gauge-preserving local transformations gives rise to the holographic Ward identities.\footnote{In fact, gauge-preserving bulk diffeomorphisms and local supersymmetry transformations cannot be considered separately since they mix. However, this mixing occurs only at asymptotically subleading orders and involves transverse derivatives on the transformation parameters \cite{usfut}. This implies that the mixing between gauge-preserving bulk diffeomorphisms and local supersymmetry transformations does not affect our results here, and so for simplicity we will treat them separately.}

\subsection{Bulk diffeomorphisms}

Infinitesimal bulk diffeomorphisms that preserve the gauge \eqref{gf-conditions} are parameterized by a vector field satisfying the differential equations 
\be
%\boxed{
\dot\x^r=0,\quad \dot\x^\m+\g^{\m\n}\pa_\n\x^r=0 \ . %}
\ee
The general solution of these equations is 
\bal
\label{weylres}
&\x^r =\s(x) \ , \\
&\x^\m=\x_o^\m(x)-\int^r dr' \g^{\m\n}(r',x)\pa_\n\s(x)\ ,
\eal
where the arbitrary functions $\s(x)$ parameterizes Weyl transformations on the boundary \cite{Imbimbo:1999bj}, while $\x_o^\m(x)$ corresponds to boundary diffeomorphisms. The transformation of the supergravity fields under Weyl transformations \eqref{weylres} is
\bal\label{weyl-transf}
&\d_\s\g_{\m\n}=\s\dot\g_{\m\n}\sim 2e^{-\frac{8}{15}U}\s\g_{\m\n}\ ,
&&\d_\s\J_\m^+=\s\dot\J_\m^+\sim \frac12e^{-\frac{8}{15}U} \J_\m^+\sigma\ ,\NO\\
% % % % % % % % % % % % % % % % % % % % % % % % % % % % % % % % % % % % %
&\d_\s\f=\s\dot\f\sim 0 \ , 
&&\d_\s\z_\f^-=\s\dot \z_\f^-\sim -\frac12e^{-\frac{8}{15}U}\z_\f^-\s \ ,\\
% % % % % % % % % % % % % % % % % % % % % % % % % % % % % % % % % % % % % 
&\d_\s b^\F=\s\dot b^\F\sim  qe^{-\frac{8}{15}U+\f}\s\ , 
&&\d_\s\z_b^-=\s\dot \z_b^-\sim -e^{-\frac{8}{15}U}\left(\frac12\z_b^- -q e^{\f}\left(\z^-_\f+\frac{8}{15}\z_U^-\right)\right)\s\ ,\NO\\
% % % % % % % % % % % % % % % % % % % % % % % % % % % % % % % % % % % % %
&\d_\s U= \s\dot U\sim  \frac58q^2e^{-\frac43 U+\f}\s\ , 
&&\d_\s\z_U^-=\s\dot \z_U^-\sim -\frac{3q}{16}e^{-\frac43U}\left(\z_b^--\frac{7q}{4}e^\f\z_\f^-\right)\s \ .\NO
\eal
These imply that the covariant sources transform as 
\bal\label{weyl-transf-sources}
&\d_\s\wt\g_{\m\n}\sim 2e^{-\frac{8}{15}U}\s\wt\g_{\m\n}\ ,
&&\d_\s\wt\J_\m^+\sim \frac12e^{-\frac{8}{15}U} \wt\J_\m^+\sigma\ ,\NO\\
% % % % % % % % % % % % % % % % % % % % % % % % % % % % % % % % % % % % %
&\d_\s\f\sim 0 \ , 
&&\d_\s\z_\f^-\sim -\frac12e^{-\frac{8}{15}U}\z_\f^-\s \ ,\NO\\
% % % % % % % % % % % % % % % % % % % % % % % % % % % % % % % % % % % % % 
&\d_\s \wt b^\F\sim q e^{-\frac{8}{15}U}\s\ , 
&&\d_\s\wt \z_b^-\sim -\frac12e^{-\frac{8}{15}U} \(\wt\z_b^- -\frac{16}{15}q e^\f\z_U^{-}\)\s \ ,\NO\\
% % % % % % % % % % % % % % % % % % % % % % % % % % % % % % % % % % % % %
&\d_\s \wt U\sim 0\ , 
&&\d_\s\wt \z_U^-\sim 0 \ .
\eal

\subsection{Local supersymmetry transformations}

The gauge fixing condition \eqref{gf-conditions} on the gravitino leads to a differential equation for the supersymmetry parameter $\e$ via eq.~\eqref{gravitinovar}, namely
\be
\left(\nabla_r+\frac16\cw\G_r\right)\e=0 \ ,
\ee
or, in gauge-fixed form and projecting out the two chiralities,  
\be %\boxed{
\dot\e^\pm\mp\frac16\cw\e^\pm=0 \ . %}
\ee
The asymptotic solutions to these equations are
\bal
&\e^+(z,x)=z^{-1/2}h(z)^{1/12}\e_0^+(x)+\co(z^4)\ ,\NO\\
&\e^-(z,x)=z^{1/2}h(z)^{-1/12}\e_0^-(x)+{\cal O}(z^4)\ ,
\eal
where the arbitrary spinors $\e_0^\pm(x)$ parameterize respectively supersymmetry and superWeyl transformations on the boundary. The transformation of the covariant sources under these transformations is as follows. 
\begin{itemize}
\item[] \hskip-0.95cm\textbf{Gravitino:} 

The transformation of the induced gravitino $\J_\m$ under supersymmetry is
\be
\d_\e\J_\m  =\(\nabla_\m +{1\over 6}{\cal W}\G_\m\)\, \e\ .
\ee
Projecting this equation on the positive chirality, which is the leading one asymptotically as follows from eq.~\eqref{f-asymptotics} and which corresponds to the covariant source of the supercurrent, we get
\be\label{dPsi}
\d_\e\J^+_\m=\del_\m\e^+ +{1\over 3}\G_\m{\cal W}\, \e^-\ ,
\ee
where we have used \eqref{BPSeqs} in order to drop a term proportional to the VEV of the stress tensor (which is subleading asymptotically).

\item[] \hskip-0.95cm\textbf{Metric:} 

The supersymmetry transformation of the vielbein $e_\m^a$ is given by
\be
\d_\e e_\m^a={1\over 2}\, \ol{\e}\, \g^a\, \J_\m+\mathrm{h.c.}\,.
\ee
From this it follows that the corresponding variation of the induced metric is
\be
\d_\e\g_{\m\n} = \ol{\e}^+\G_{(\m}\J^+_{\n)}+\ol{\e}^-\G_{(\m}\J^-_{\n)}+\mathrm{h.c.}\ ,
\ee
where the symmetrization is done with a factor of 1/2. Dropping the term proportional to $\J^-_\m$ that is related to the one-point function of the supercurrent and is asymptotically subleading, we obtain  
\be
\d_\e\g_{\m\n} = \ol{\e}^+\G_{(\m}\J^+_{\n)}+\mathrm{h.c.}\ .
\ee

\item[] \hskip-0.95cm\textbf{Hypermultiplet sector:} 

The transformation of the fields in the hypermultiplet is
\begin{align}
\label{}
&\d_\e\f = {i\over 2}\ol{\e}^+\z^{-}_\f+\mathrm{h.c.}\ ,
&&\d_\e\z^{-}_\f=-{i\over 2}\G^z\del_z\f~\e^-\sim 0\ ,\\
% % % % % % % % % % % % % % % % % % % % % % % % % % % % % % % % % % % % % %
&\d_\e b^\F = {i\over 2}\ol{\e}^+\z^{-}_b+\mathrm{h.c.}\ , 
&&\d_\e\z^{-}_b=-{i\over 2}\(\G^z\del_z b^\F+e^{-{4\over 15}(2U-3V)+\f}q\)\e^-\sim -iqe^{-{8\over 15}U+\f}\e^-\ ,\NO\\
% % % % % % % % % % % % % % % % % % % % % % % % % % % % % % % % % % % % % %
&\d_\e U=\frac i2\ol\e^+\z_U^-+\mathrm{h.c.}\ ,
&&\d_\e\z_U^-=-\frac i2\left(\G^z\pa_z U-\pa_U\cw\right)\e^-\sim i\pa_U\cw\e^-\sim -i\frac{q^2}{3}e^{-\frac43 U+\f}\e^-\ .\NO
\end{align}

\end{itemize}

Combining these results, we deduce that the covariant sources transform under $\e^\pm$ as 
\begin{align}
\label{sources-e+}
&\d_{\e^+}\wt\g_{\m\n}\sim \ol{\e}^+\wt \G_{(\m}\wt \J^+_{\n)}+\mathrm{h.c.}\ , &&\d_{\e^+}\wt\J^+_\m\sim e^{-\frac{2}{15}U}\pa_\m\e^+\ ,\NO\\
% % % % % % % % % % % % % % % % % % % % % % % % % % % % % % % % % % % % % % 
&\d_{\e^+}\f \sim {i\over 2}\ol{\e}^+\z^{-}_\f+\mathrm{h.c.}\ ,
&&\d_{\e^+}\z^{-}_\f\sim 0\ ,\NO\\
% % % % % % % % % % % % % % % % % % % % % % % % % % % % % % % % % % % % % %
&\d_{\e^+} \wt b^\F \sim {i\over 2}\ol{\e}^+\wt\z^{-}_b+\mathrm{h.c.}\ , 
&&\d_{\e^+}\wt\z^{-}_b \sim 0\ ,\NO\\
% % % % % % % % % % % % % % % % % % % % % % % % % % % % % % % % % % % % % %
&\d_{\e^+} \wt U\sim {i\over 2}\ol{\e}^+\wt\z^{-}_U+\mathrm{h.c.}\ ,
&&\d_{\e^+}\wt \z_U^-\sim 0\ .
\end{align}
where $\wt\G_\m=\wt e_\m^a\g_a=e^{-\frac{2}{15}U} e_\m^a\g_a$, and 
\begin{align}
\label{sources-e-}
&\d_{\e^-}\wt\g_{\m\n}\sim 0\ , &&\d_{\e^-}\wt\J^+_\m\sim e^{-\frac{8}{15}U}\wt\G_\m \e^-\ ,\NO\\
% % % % % % % % % % % % % % % % % % % % % % % % % % % % % % % % % % % % % % 
&\d_{\e^-}\f \sim 0\ ,
&&\d_{\e^-}\z^{-}_\f\sim 0\ ,\NO\\
% % % % % % % % % % % % % % % % % % % % % % % % % % % % % % % % % % % % % %
&\d_{\e^-} \wt b^\F \sim 0\ , 
&&\d_{\e^-}\wt\z^{-}_b \sim -iqe^{-\frac{8}{15}U}\e^-\ ,\NO\\
% % % % % % % % % % % % % % % % % % % % % % % % % % % % % % % % % % % % % %
&\d_{\e^-} \wt U\sim 0\ ,
&&\d_{\e^-}\wt \z_U^-\sim -i\frac{7}{30} q^2 e^{-\frac43U+\f}\e^-\ .
\end{align}

\newpage

\bibliographystyle{plainnat}

\end{document}